%% file: main.tex
\documentclass[notitlepage, 12pt]{article}

\usepackage{amssymb,amsmath,amsfonts,eurosym,geometry,ulem,graphicx,color,setspace,sectsty,comment,footmisc, achicago,pdflscape,array,hyperref,subcaption,threeparttable, rotating, chngcntr, pifont, mathtools, longtable, threeparttablex,cite}

\usepackage{booktabs}
\usepackage{longtable}
\usepackage{adjustbox}

\usepackage[font=small]{caption}
\usepackage{rotating}

\renewcommand{\cite}{\citeasnoun}

\def\bsk{\bigskip}                      
\def\msk{\medskip}
\def\ssk{\smallskip}

\begin{document}
\begin{titlepage}

\title{\textbf{Credit Scores: Performance and Equity}\thanks{ We are grateful to Dokyun Lee, Sera Linardi, Yildiray Yildirim, Albert Zelevev  and  many seminar participants  for useful comments and suggestions. This research was supported in part by the National Science Foundation under Grant No. SES 1824321. This research was also supported in part by the University of Pittsburgh Center for Research Computing through the resources provided. Correspondence to: stefania.albanesi@gmail.com.} }

\author{Stefania Albanesi, University of Miami, NBER and CEPR 
\and
Domonkos F. Vamossy, University of Pittsburgh
}

\maketitle
\thispagestyle{empty}

\begin{abstract}
Credit scores are critical for allocating consumer debt in the United States, yet little evidence is available on their performance. We benchmark a widely used credit score against a machine learning model of consumer default and find significant misclassification of borrowers, especially those with low scores. Our model improves predictive accuracy for young, low-income, and minority groups due to its superior performance with low quality data, resulting in a gain in standing for these populations. Our findings suggest that improving credit scoring performance could lead to more equitable access to credit.

\msk
\bsk


\bsk
\msk

\end{abstract}

\end{titlepage}
%
%
%

\begin{doublespacing}
\setcounter{page}{1}
\input{Sections/introduction}
\input{Sections/data}

\input{Sections/model}

\input{Sections/credit_score}

\input{Sections/vulnerable}

\input{Sections/conclusion}

\clearpage

\end{doublespacing}

\clearpage 
\begin{singlespacing}
\bibliography{cites}
\bibliographystyle{achicago}
\end{singlespacing}

\clearpage

\input{Sections/appendix}

\end{document}

%% file: Sections/introduction.tex
\section{Introduction}\label{sec:intro}





Credit scores are key in the allocation of consumer credit in the United States. Lenders report the status of consumer loans to three credit bureaus-- private corporations that collect administrative data on most consumer loans.\footnote{Credit reporting started in a very basic form in the 1880s and developed to a comprehensive nationwide system in the 1930s (\citeN{lauer_2017}). The Fair Credit Reporting Act of 1970 systematized the information that can be collected in credit reports and determined which financial entities are subject to mandatory reporting \citeN{hunt2005century}.} 
Using proprietary models, these credit scores rank borrowers based on their likelihood of future default, providing a single, standardized measure of individual risk. The Fair Isaac Corporation introduced the first credit score in 1958 and launched the widely used FICO score in 1989. The FICO score and the VantageScore, introduced in 2006, are the most commonly used.\footnote{There are also several other scores that are product specific, for example, for vehicle loans or credit cards (see \citeN{CFPB_2012}). Credit scoring models are also updated regularly and change over time. More information on credit scores is reported in Section \ref{sec:credit_score} and Appendix \ref{app:score}. }
When consumers apply for a new loan, lenders check their credit score. While the credit score may not be the only input in the underwriting decision, the amount of credit and the interest rate mostly depend on the consumer's credit score, even for secured debts, such as mortgages and vehicle loans (\citeN{edelberg2006risk}). The emergence of algorithmic underwriting of consumer loans based on credit scores  is seen as a key factor in the growth of consumer debt in the past three decades (\citeN{avery2003overview}).

Despite their ubiquitous use in the financial industry, there is very little information on credit scores with no performance metrics publicly available. Emerging evidence suggests that  currently used credit scores have severe limitations. The Consumer Financial Protection Bureau estimates that 11\% of consumers are unscored, and therefore excluded from conventional credit markets, with these borrowers concentrated among young, low-income, and minority populations (\citeN{CFPB_2016_unscored}). Additionally, \citeN{albanesi2022credit} show that during the 2007-2009 housing crisis, there was a marked rise in mortgage delinquencies among high credit score borrowers, suggesting that credit scoring models at the time did not accurately reflect the probability of default for these borrowers. 

We assess the performance of a widely used credit score by developing an alternative scoring model based on machine learning with substantially better predictive performance. We find that credit scores misclassify 41\% of consumers by placing them in a risk category that does not align with their actual default probability. The misclassification is more severe for consumers with low credit scores. For example, 47\% of Subprime and 70\% of Near Prime borrowers are misclassified, while only 26\% of Superprime borrowers are. We show that credit score performance is worse for young, low-income, and minority borrowers and our model improves performance most for these borrowers. Our model also improves the ranking of these marginalized borrowers, which could contribute to a fairer and more equitable distribution of consumer credit.

The machine learning model of consumer default that we deploy to benchmark credit score performance uses the same information as standard credit scoring models and is designed for environments with high-dimensional data and complex non-linear patterns of interaction among variables.\footnote{For excellent reviews of how machine learning can be applied in economics, see \citeN{mullainathan2017machine} and \citeN{athey2019machine}.}
Our model targets the same default outcome as conventional credit scoring models: a 90+ days delinquency  within the subsequent eight quarters, and strictly uses only the information in credit reports that is permitted under current legislation.
Since credit scores provide only an ordinal ranking of consumers based on their default risk, we use only the ordinal ranking of consumers with respect to their predicted probability of default based on our model. 

Our model performs significantly better than conventional credit scores. The average AUC score for the credit score is about 85\%, but it drops notably during the 2007-2009 crisis, while the average AUC score for our model is approximately 91\% and stable over time. Most importantly, the credit score generates significant disparities between the implied predicted probability of default and the realized default rate for large groups of borrowers, particularly at the low end of the credit score distribution. We show that, among borrowers with a Subprime credit score, who comprise 21\% of the population, 22\% display default behavior consistent with Near Prime borrowers, and 15\% display default behavior consistent with Deep Subprime. The realized default rates for Deep Subprime, Subprime, and Near Prime borrowers are 68\%, 44\%, and 22\%, respectively. This suggests that the credit score is unable to differentiate between borrowers with substantially different default risks. By contrast, the discrepancy between predicted and realized default rates for our model is never more than five percentage points within any risk category. Our analysis points to severe limitations of conventional credit scores in their ability to differentiate consumers by default risk.

We use interpretability techniques to identify which factors are associated with variation in consumer default ranking for our model, and compare these to the credit score. By law, credit scoring companies have to reveal the four most important factors driving credit score variation, which are reported to be amounts owed, credit mix, incidence of new credit, and length of the credit history. Our model places more weight on amounts owed, which explains 49\% of the variation in our model implied rankings, and only 30\% of the variation in credit scores. By contrast, credit mix and the incidence of new credit, which are viewed as indicators of credit demand, each only account for 5\% of the variation in rankings for our model, while they each account for 10\% of the variation in credit scores. Additionally, the length of the credit history only accounts for 8\% of variation in rankings for our model, but 15\% of the variation in credit scores. 


One standing concern with the adoption of machine learning-based scoring models is that a more sophisticated statistical technology might exacerbate disparities in access to credit for disadvantaged consumers such as young, low-income, or minority borrowers (see \citeN{fuster2018predictably}). We show that, on the contrary, our model provides a more favorable risk assessment to young, low-income, and in most cases minority borrowers, particularly for those who do not default. This result likely stems from the property that credit demand factors and length of the credit history have a sizable negative impact on the conventional credit score, while they play a lesser role in default predictions made by our model. Young, minority, and low-income borrowers, who often have short credit histories and high demand for credit, are more at risk of being placed lower in the ranking by the credit score.
Additionally, we show that our machine learning model improves performance relative to the credit score more for young, low-income, and minority consumers compared to the rest of the population, and that this result is a function of its better ability to deliver accurate predictions when confronted with low quality data, such as thin files, low credit mix and a history of default, attributes that are prevalent among these traditionally marginalized populations. Contrary to \citeN{blattner2021costly}, these findings suggest that a credit scoring algorithm based on machine learning can improve performance in the presence of data bias, benefitting vulnerable consumers.



Our analysis contributes to the literature on credit scores and disparities in credit access. Credit scores are viewed as a key tool for improving consumer credit market efficiency (\citeN{chatterjee2016theory}), yet this role is based on the premise that they are an unbiased if imperfect signal of the variation in default risk amongst consumers. However, we show that widely used credit scores place too much weight on factors that are not strongly associated with default and misclassify a large fraction of consumers. Additionally, the statistical performance of the credit score is worse for populations historically marginalized on credit markets and these consumers would gain in standing if ranked by a better performing model. This implies that our default predictions could help improve credit allocation in a way that benefits both lenders, in the form of lower losses from default, and borrowers, in the form of more access to credit for vulnerable populations. Our results also speak to the perils associated with using conventional credit scores outside the consumer credit sphere. As it is well known, credit scores are used to screen job applicants, in insurance applications, and a variety of additional settings. Economic theory would suggest that this is helpful as long as the credit score provides information correlated with hidden characteristics of interest to the party using the score (\citeN{Corbae_Glover_2018}). However, as we show, conventional credit scores misclassify borrowers by a substantial degree based on their default risk and are biased against young, low-income, and, in most cases, minority borrowers, which implies that they may not be accurate and may not include appropriate information or use adequate methodologies. The expanding use of credit scores could amplify the economic disparities resulting from these limitations.


The remainder of this paper is structured as follows. Section \ref{sec:data} describes our data. Section \ref{sec:model} describes our prediction problem and our model. Section \ref{sec:credit_score} compares our model to conventional credit scores. Section \ref{sec:vulnerable} investigates the trade-off between performance and equity.

%% file: Sections/data.tex
\section{Data}\label{sec:data}

We use anonymized credit file data from the Experian credit bureau. The data is quarterly. It starts in 2004Q1 and ends in 2015Q4. The data comprises over 200 variables for a nationally representative panel of 1 million households, constructed with a random draw from the universe of borrowers with an Experian credit report. The data covers credit cards and other revolving credit, auto loans, installment loans, business loans, first and second mortgages, home equity lines of credit, student loans, and collections. There is information on the number of trades for each type of loan, the outstanding balance and available credit, the monthly payment, hard inquiries, and whether any of the accounts are in a state of delinquency. All balances are adjusted for joint accounts to avoid double counting. We also have each borrower's credit score for each quarter in the sample. Because this is data drawn from credit reports, we do not know the gender, marital status, or any other demographic characteristic, though we know a borrower's address at the zip code level. We also do not have any information on asset holdings. The data also includes an estimate of individual and household labor income based on IRS data. 

\input{Tables/desc_stats}

Table \ref{tab:desc_stats} reports basic demographic information on our sample, including age, household income, credit score, and incidence of default, defined as the fraction of households who report 90 or more days past due delinquency on any trade, excluding collections. This will be our baseline definition of default, as it is the outcome targeted by credit scoring models.

%% file: Tables/desc_stats.tex
\begin{table}[htbp]\centering
\begin{threeparttable}
\caption{Descriptive Statistics}\label{tab:desc_stats}
\small
\begin{tabular}[l]{p{2in}p{0.5in}p{0.5in}p{0.25in}p{0.25in}p{0.25in}p{0.25in}p{0.25in}p{0.2in}}
Feature & Mean & Std. Dev. & Min &  25\% & 50\% & 75\% & Max   \\ \hline 
&&&&&&& \\[\dimexpr-\normalbaselineskip+2pt]
Age & 45.530 & 16.751 & 18.0 & 31.0 & 45.0 & 58.0 & 84.0 \\
Household Income (Imputed) & 78.331 & 55.092 & 15.0 & 43.0 & 64.0 & 91.0 & 316.0 \\
Credit Score & 679.317 & 108.475 & 300.0 & 595.0 & 691.0 & 776.0 & 839.0 \\
Credit History (Months) & 197.672 & 129.669 & 0.0 & 95.0 & 175.0 & 274.0 & 988.0 \\
$\mathbb{I}_{\{\text{90+ DPD Debt}\}}$ & 0.078 & 0.269 & 0.0 & 0.0 & 0.0 & 0.0 & 1.0 \\
$\mathbb{I}_{\{\text{90+ DPD Debt, next 8 Quarters}\}}$ & 0.184 & 0.387 & 0.0 & 0.0 & 0.0 & 0.0 & 1.0 \\
\hline  \hline 
\end{tabular}
\begin{tablenotes}
\footnotesize
\item Notes:  Household income is in USD thousands, winsorized at the 99th percentile. Source: Authors' calculations based on Experian Data.
\end{tablenotes}
\end{threeparttable}
\end{table}

%% file: Sections/model.tex
\section{Model}\label{sec:model}

Credit scores are an ordinal measure that ranks consumers by the probability that they will default over a certain future horizon based on information in a borrowers' credit report. The definition of default used in this context is 90 days or more past due on any debt, and the horizon is 8 quarters.
Given that credit scoring models are proprietary, little is known about how the models are constructed. The Fair Credit Reporting Act of 1970 and the Equal Opportunity in Credit Access Act of 1974, and their subsequent additions, regulate credit scores and in particular determine which information can be included and must be excluded in credit scoring models. Such models can incorporate only information in a borrower's credit report, except for age and location. These restrictions are intended to prevent discrimination by age and factors related to location, such as race.\footnote{Credit scoring models are also restricted by law from using information on race, color, gender, religion, marital status, salary, occupation, title, employer, employment history, nationality.} The law also mandates that entities that provide credit scores make public the four most important factors affecting scores. In marketing information, these are reported to be payment history, which is stated to explain about 35\% of variation in credit scores, followed by amounts owed, length of credit history, new credit and credit mix, explaining 30\%, 15\%, 10\% and 10\% of the variation in credit scores respectively.\footnote{ See for example \url{https://www.myfico.com/credit-education/whats-in-your-credit-score}.} 
There is no mandated requirement to reveal any performance information on credit scores.\footnote{The FACT Act of 2003 required the Federal Reserve Board to study some basic performance properties of credit cores. The findings are summarized in \citeN{reserve2007report}.}

To evaluate the performance of credit scores, we develop a model to predict consumer default based on machine learning. We then use the ordinal ranking of consumers based on their predicted probability of default as a credit score alternative and compare the performance of conventional credit scores to our model.

Predicting consumer default maps well into a supervised learning framework, one of the most widely used techniques in machine learning. In supervised learning, a learner takes in pairs of input/output data. The input data, typically a vector, represent preidentified attributes, also known as features, used to determine the output value. The supervised learning problem is referred to as a ``regression problem" when the output is continuous and a ``classification problem" when the output is discrete. Once the learner is presented with input/output data, its task is to find a function that maps the input vectors to the output values. 
The goal of supervised learning is to find a function that generalizes beyond the training set to forecast out-of-sample outcomes correctly. Adopting this machine-learning methodology, we build a model that predicts defaults for individual consumers. %

We now formalize our prediction problem. We define default as a 90+ days delinquency on any debt in the subsequent 8 quarters, the outcome targeted by conventional credit scoring models. We adopt a discrete-time formulation for periods 0,1,..., T, each corresponding to a quarter. We let the variable $D_t^i$ prescribe the state at time $t$ for individual $i$ with D $\subset \mathbb{N}$ denoting the set of states. We define $D_1^i=1$ if a consumer is 90+ days past due on any trade and $D_1^i=0$ otherwise. Consumers will transition between these two states quarter to quarter. 
Our target outcome corresponds to:
\begin{equation}
Y_t^i=\left\{ \begin{array}{cl}
0 & \textrm{if }\sum_{n=t}^{t+7} D_n^i = 0\\
1 & \textrm{otherwise}\\
\end{array}\right.
\end{equation}

We posit that the dynamics of this process are influenced by a vector of explanatory variables $X_{t-1}^i$ $\in \mathbb{R}^{d_X}$, which includes the previous realization of the state $D_{t-1}^i$. In our empirical implementation, $X_{t-1}^i$ represents the set of features, which we will discuss below.
 We fix a probability space $(\Omega,\mathcal{F},\mathbb{P})$ and an information filtration $(\mathcal{I}_t)_{(t=0,1,...,T)}$. Then, we specify a probability transition function $f_\theta : \mathbb{R}^{d_X} \rightarrow [0,1]$ satisfying
\begin{equation}\label{eq:trans}
 \mathbb{P}[Y_t^i = y| \mathcal{I}_{t-1}] = f_\theta(X_{t-1}^i), y \in D 
\end{equation}
where $\theta$ is a vector of parameters to be estimated. Equation \ref{eq:trans} gives the marginal conditional probability for the transition of individual $i$ from state $D_{t-1}^i$ at time $t-1$ to state $y$ at time $t$ given the explanatory variables $X_{t-1}^i$.\footnote{The state $ y$ encompasses realizations of the state between time $t$ and $t+7$.} 
The vector output of the function $g$ is a probability distribution on $D$. 
Our model outputs a continuous variable between 0 and 1 that can be interpreted as an estimate of the probability of default for a particular borrower at a given time, given input variables from their credit reports. 

Equation \ref{eq:trans} defines a theoretical transition matrix for the default outcome we consider. Its empirical counterpart is reported in Table \ref{tab:transition}. The fraction of consumers in our sample in default according to this definition is 18.4\%. The default and no-default state are very persistent, with a 85\% of consumers in default and 87\% of consumers not in default remaining in the same state in the subsequent quarter.

\input{Tables/default_transitions}

We adopt a deep learning model, as models in this class display high performance in settings with high volume data and  multidimensional non-linear interactions.\footnote{\citeN{sirignano} and \citeN{fuster2022predictably} apply these models to mortgage default risk.} Specifically, we model the transition function $f_\theta$ with a hybrid deep neural network/gradient boosting model, which combines the predictions of a deep neural network and an extreme gradient boosting model. This hybrid approach is particularly well-suited to our prediction problem. The deep neural network excels at learning intricate patterns, while the gradient boosting model improves its predictive accuracy by correcting errors from previous iterations, making this combination the best-performing model in this class for our specific prediction problem and data (\citeN{albanesi_vamossy_NBER_2019}). Appendix \ref{app:model_details} explains each of the component models and their properties and the rationale for combining them.\footnote{\citeN{albanesi_vamossy_NBER_2019} extensively discusses alternative machine learning models and their performance properties for this prediction problem. Additionally, it explains the rationale for adopting this class of models.} 

%
%
%
%

We implement our approach as follows. Given the 8-quarter horizon for the prediction problem, we allow for fully out-of-sample performance evaluation by separating our training/validation and testing data by 8 quarters. Additionally, we only use one quarter of data to train and validate our model. That is, we train and validate our model on quarter $t$ data and then test it on quarter $t+8$ data.\footnote{This type of temporal cross-validation approach within the credit scoring context was also adopted by \citeN{khandani}.} For example, default predictions for quarter 2008Q2 will be generated with a model trained on 2006Q2 data. This allows us to provide an out of sample default prediction for every consumer with non-empty credit record in a given quarter for 2006Q1-2016Q2. We select a broad set of 79 features, listed in  Table \ref{tab:features}, in order to maximize the number of consumers with a prediction and cover the factors that are listed by conventional credit scoring models, such as credit mix, length of credit history, utilization, and new credit. Appendix \ref{app:model_details} presents detailed performance metrics for our model. 


\begin{table}
\footnotesize
\begin{center}
\begin{tabular}{p{1.5cm} p{2.cm}p {2.cm} p{2.cm} } \hline 
Data & Training & Validation & Testing \\ \hline
Purpose & fitting &hyperparameter tuning & out of sample evaluation \\
&&& \\
Time & $t$ & $t$ & $t+8$ \\
& \multicolumn{2}{c}{ random 80/20 split } & \\ \hline
\end{tabular}
\end{center}
\end{table}


%% file: Tables/default_transitions.tex
\begin{table}\caption{$Y_t^i$ Frequency and Transitions}\label{tab:transition}
\begin{center}
\begin{threeparttable}
\small
\begin{tabular}{ll|rll} \hline \hline 
Frequency &  & Transition probability & No default & Default \\ \hline
& & & & \\[\dimexpr-\normalbaselineskip+3pt] 
Default & 0.184 & No default & 0.8729 & 0.1271 \\
No default & 0.816 & Default & 0.1460 & 0.8540 \\ \hline \hline
\end{tabular}

\begin{tablenotes}
\footnotesize
\item Notes: Average default rates and average quarterly transition frequency 2006Q1-2016Q2. Source: Authors' calculations based on Experian Data.
\end{tablenotes}
\end{threeparttable}
    
\end{center}
\end{table}

%% file: Sections/credit_score.tex
\section{Comparison with Credit Score}\label{sec:credit_score}

We now compare the performance of our model to a widely used conventional credit score. Our goal is to provide a conceptual assessment of credit score performance based on its economic role in consumer credit markets. We are not able to provide a comprehensive statistical comparison for two reasons. First, we do not know the precise attributes or the particular time frame used to build credit scoring models, as the law only requires disclosing basic features of those models (see Section \ref{app:feature_attribution}). Second, there might be additional constraints that credit scoring companies impose on their models, driven by monotonicity and palatability requirements.\footnote{See \url{https://www.fico.com/blogs/trusted-ai-challenge-monotonicity-and-palatability} for a discussion.} These  requirements are imposed on credit scoring models to enhance transparency and facilitate equitable treatment of consumers. We apply explainability techniques to make our model transparent, and as we will show in Section \ref{sec:vulnerable}, ranking consumers based on our model advances the standing of traditionally marginalized consumers. So, even while not explicitly imposing these constraints, our model satisfies them indirectly.

Since the credit score is an ordinal ranking of individuals with respect to default risk, we only use the ordinal ranking of consumers with respect to the probability of default implied by our model for our comparison with the credit score. Specifically, when comparing the ranking of consumers implied by our model and by the credit score, we proceed as follows. We utilize a credit score that is a whole number, ranging from 300 to 850, with higher values corresponding to lower default risk, which we segment into percentiles. Since  our model's output, $\hat{Y}(X)$, is a probability of default, we rank consumers by: 1 - $\hat{Y}(X)$, segmenting the corresponding distribution into 100 equal-sized bins (percentiles). 

Our analysis proceeds in three steps. We first evaluate the statistical performance of the credit score and our model using standard performance metrics. We then evaluate the ability of the credit score to place consumers in the standard risk profiles used in the industry, such as Subprime and Prime, and show that the credit score misclassifies a substantial fraction of borrowers, especially at the bottom of the credit score distribution. We also examine the factors that lead to different default predictions for the credit score and our model. Finally, we study the impact on vulnerable populations by age, income and race.


\subsection{Statistical Performance}\label{subsec: ranking}

\begin{figure}[!h]
	\centering
		\includegraphics[scale=0.5]{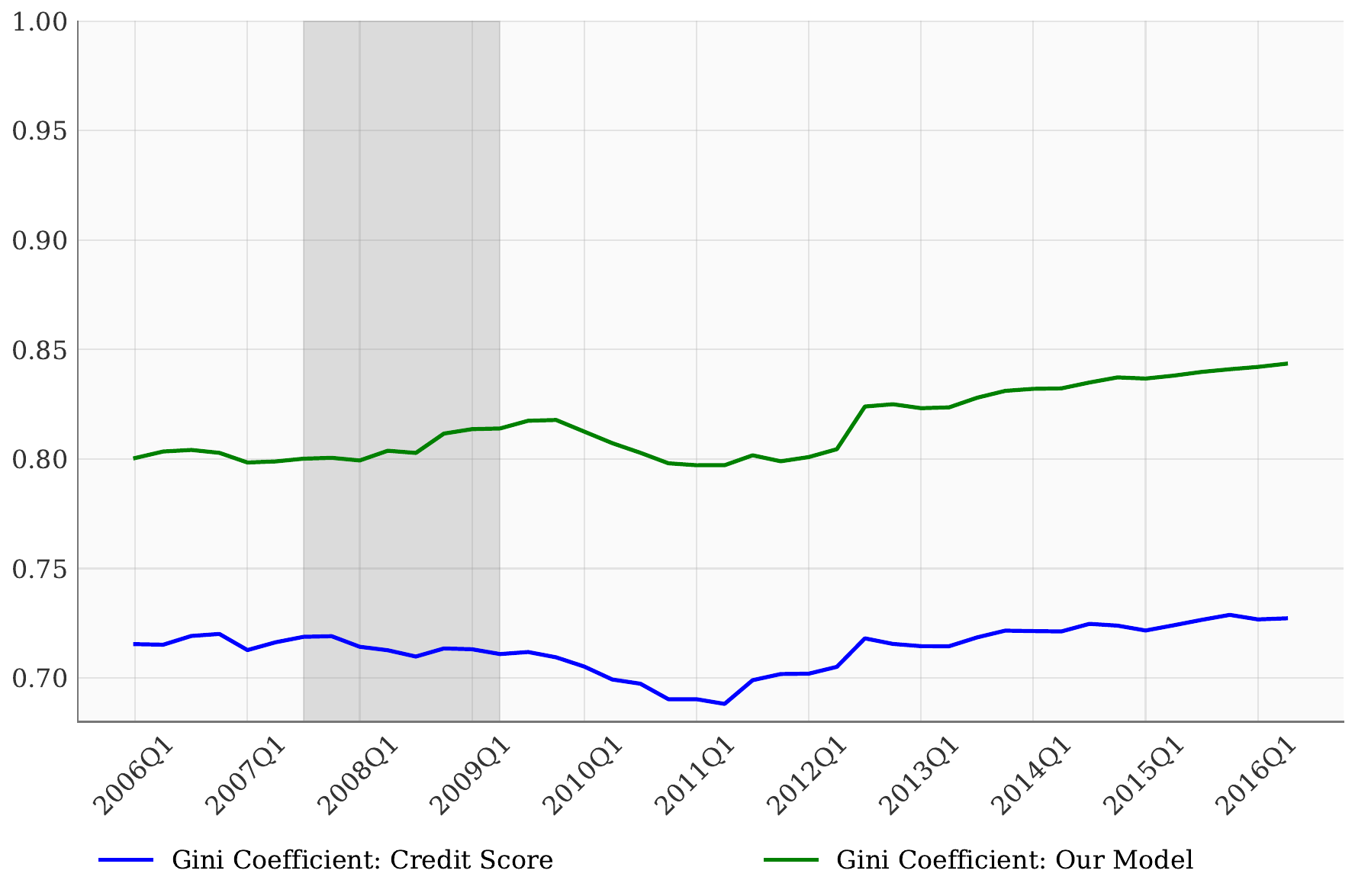} 
	
\caption{Gini Coefficient, Credit Score and Prediction Model}\label{fig:rank_gini_comp}
		\begin{flushleft}
		 \footnotesize{Notes: Gini coefficients for the credit score and our model's by quarter. Source: Authors' calculations based on Experian data.}
		\end{flushleft}
\end{figure}

A common way to measure performance for conventional credit scoring models is the Gini coefficient, which measures the dispersion of the credit score distribution and, therefore, its ability to separate borrowers by their default risk. The Gini coefficient is related to a key performance metric for machine learning algorithms, the AUC score, with $Gini=2*AUC-1$ so that we can compare the credit score performance to our model along this dimension. Figure \ref{fig:rank_gini_comp} plots the Gini coefficient for the credit score and our model prediction over time. The Gini coefficient for our model is about 0.8 between 2006Q1 and 2008Q3 and then rises to 0.84 for the rest of the sample. For the credit score, the Gini coefficient is above 0.7 until 2011Q1, when it drops to approximately 0.69, then slowly recovers to 0.72, suggesting a drop in credit score performance in the aftermath of the Great Recession. Appendix \ref{app:score} presents similar results for the Spearman and Kendall rank correlations. This analysis highlights the superior separating power of our model compared to the conventional credit score, particularly during periods of economic instability.


\subsection{Performance by Risk Profile}\label{app:prime_etc}

Credit scores are used to place borrowers into five industry defined risk profiles-- Deep Subprime, Subprime, Near Prime, Prime, and Super Prime-- that are critical for the determination of credit limits and interest rates on consumer loans. We now examine the performance of the credit score in relation to these risk profiles.

The credit score threshold levels that correspond to the risk profiles in descending order of default risk for the credit score are reported in Table \ref{tab:risk_def}. As shown in the table, these categories account for respectively, 6\%, 20.1\%, 14.0\%, 36.3\% and 23.6\% of all borrowers. The thresholds that define these categories can be restated in percentiles of the credit score distribution, which are reported in the second row of the table. Based on this classification, we can create five corresponding risk profiles for the predicted probability of default based on our model, with the lowest category containing the 6\% of borrowers with the highest predicted default risk, corresponding to Deep Subprime, all the way to the highest category containing the 23.6\% of all borrowers with the lowest predicted default risk, corresponding to Super Prime. We then calculate where borrowers in each credit score risk profile are placed by our model. 

\input{Tables/risk_def}

The results are displayed in Table \ref{tab:risk_comp}, with items on the diagonal showing the fraction of borrowers that our model would place in the same risk profile as the credit score, and off-diagonal items showing the fraction of borrowers in each credit score risk profile who are misclassified according to our model. The last column shows the overall percentage of borrowers who are misclassified by the credit score for each credit score risk profile.

The findings suggest that credit scores misclassify a significant portion of borrowers, especially those with lower credit scores. According to our model, only 45\% of Deep Subprime borrowers would remain in that category, while 44\% would be reclassified as Subprime, 9\% as Near Prime, and 2\% as Prime. For Subprime borrowers, our model places 53\% in the Subprime category, with 15\% reclassified as Deep Subprime and 22\% as Near Prime. Among Near Prime borrowers, our model categorizes 38\% as Subprime, 28\% as Prime, and only 30\% as Near Prime. In the Prime category, 67\% of borrowers remain in that profile, while 13\% are reclassified as Near Prime and 16\% as Super Prime. Finally, for Super Prime borrowers, our model places 74\% in that category, with 26\% reclassified as Prime. The credit score classifies between 26\% and 70\% of borrowers into the wrong risk category, with higher misclassification rates for those with lower credit scores.
 
\input{Tables/risk_comp}

The differences in assigned risk profiles between the credit score and our model are significant, reflecting substantial variations in default risk, as illustrated in Table \ref{tab:cs_comp}. For example, Subprime borrowers have an average default rate of 44\%. However, borrowers who are Subprime based on the credit score but Near Prime based on our model have a 17\% realized default rate, whereas those who are Subprime according to our model have a 48\% default rate. Similarly, borrowers with Prime credit score have a default rate of 6\% overall. However, among them, those who have a Near Prime profile based on our model have a 13\% realized default rate, whereas those who are Prime according to our model have a 4\% realized default rate. This suggests that the credit score is unable to distinguish borrowers with very different risk profiles, whose default rates often vary by more than 10 percentage points. This is a severe limitation, as the primary goal of the credit score is to stratify borrowers based on their default risk. As we will document in Section \ref{sec:value_added_consumers}, being placed in an incorrect risk category has severe implications for consumers in terms of access to credit. 

Table \ref{tab:cs_comp} also reports our model-based prediction for the default rate of borrowers in each credit score/model risk profile. Our model-predicted default rate is always less than 5 percentage points away from the realized default rate for each risk profile. This confirms that our model's statistical performance is high at every point of the default risk distribution. Our model outperforms the conventional credit score, particularly in accurately categorizing borrowers with lower credit scores, who are traditionally more prone to misclassification. This superior performance underscores the potential for our model to improve credit risk assessment and reduce misclassification, particularly for vulnerable populations.

 \input{Tables/risk_diff_new}

\subsection{Feature Attribution}\label{app:feature_attribution}

We now examine the possible factors leading to differences in predictions of default risk ranking for our model and the credit score. We base our analysis on public information from the FICO and Vantage Score consumer websites, that cite, following the legislative requirements, the most important factors driving the credit score distribution. Credit scoring companies are required to list at least four. We  group the features we include in our model into categories to correspond to these same factors and aggregated the absolute SHAP values\footnote{For more details on SHAP, see \citeN{shap}. } for each instance across each category for our testing dataset across all periods.\footnote{Credit scoring companies do not need to reveal which specific features enter the factors they cite as driving credit score variation. For our model, we made the groupings based on the nature of each of the features.} 

Credit scoring companies cite payment history, amounts owed, length of credit history, credit mix, and new credit as the most important factors driving credit score variation. Their stated contribution towards credit scores is reported in Table \ref{tab:cs_vs_pp}. Payment history explains 35\% of the credit score variation and amounts owed 30\%. Length of credit history explains 15\% of the credit score variation, while credit mix and new credit each explain 10\% of the variation.\footnote{See \url{https://www.myfico.com/credit-education/whats-in-your-credit-score}. Payment history measures how a borrower has paid their accounts over the length of their credit history. Amounts owed includes five factors: amounts all on all accounts and on different types of accounts, accounts with balances, credit utilization ratio on existing accounts and remaining payments on installment debt. Length of credit history measures how long credit accounts have been open, the age of the newest account, and an average age of all accounts, as well as how long specific credit accounts have been open and how long it has been since accounts have been used. Credit mix refers to the number of different products, such as credit cards or mortgages, with a wider mix reported to marginally improve the credit score. New credit measures the number of new accounts and the number of inquiries in the last 12 months.} 

We report the feature attribution statistics for our model in Table \ref{tab:cs_vs_pp}. The most striking difference in feature attribution for our model in comparison to the credit score is that Amounts Owed explains 50\% of the variation in default risk rankings, compared to only 30\% for the credit score. Also, our model gives lower weight to Length of Credit History, which only explains 9\% of the variation in rankings compared to 15\% for the credit score, and New Credit, which accounts for only 3\% of variation in the model's rankings compared to 10\% for the credit score. The high weight placed by credit scores on length of credit history and new credit may explain why credit scores tend to misclassify younger and low-income borrowers to a greater degree, as we will document in the next section.

\input{Tables/feature_attribution_differences}

\subsection{Costs of Risk Profile Misclassification}\label{sec:value_added_consumers}

We have shown that the credit score's lower statistical performance leads to risk profile misclassification for a large fraction of consumers concentrated primarily at the low end of the credit score distribution. What are the costs of such risk profile misclassification? We provide an initial answer to this question by analyzing how access to credit varies as a function of credit scores for credit cards and mortgages, the most common types of consumer loans.\footnote{Since we use data from credit reports, we do not have access to interest rate information.}

Table \ref{tab:misclass_desc} in Appendix \ref{app:costs} reports summary statistics on indicators of credit demand and supply by credit score risk profile for the period 2006Q1-2016Q2. We report both the dollar values and the ratio to household income on average for each credit score risk profile. Credit limits on credit cards vary from \$7,032 on average for Deep Subprime borrowers to \$44,711 for Superprime borrowers. Mortgage balances average to \$ 172,753 for Subprime borrowers, rising to \$ 208,992 for Prime borrowers. Additionally, though demand for credit, as approximated by inquiries, is decreasing in the credit score, success at obtaining credit increases with the credit score. For example, only 6.1\% of Subprime borrowers who place a credit inquiry originate a new credit card account, while 13.6\% of Prime borrowers do. The corresponding values for mortgages are 3.9\% and 10.6\%.

These values are population averages affected by variation in characteristics, such as age and place of residence,  within a given credit score profile. To control for these factors, we estimate a regression of various credit demand and credit supply outcomes, where the main explanatory variable is the credit score risk profile, controlling for age, zip code fixed effects and quarter time effects interacted with an indicator for the credit score risk profile. The details of the regression specification and the full estimation results are reported in Appendix \ref{app:costs}. Table \ref{tab:crr_reg_summary} reports the average age-adjusted credit score risk profile effects for the entire sample period, estimated from these regression. Only 38\% of Subprime borrowers have a credit card, whereas 80\% of Prime borrowers do. This despite the fact that 11\% of Subprime borrowers show a credit card inquiry in the previous quarter, whereas 8\% of Prime borrowers do. Only 17\% of Subprime borrowers who placed a credit card inquiry at quarter $t$ or $t-1$ display a credit card origination at $t$, while 34\% of Prime borrowers do. Additionally, the average credit card balances for Subprime borrowers are \$7,730, while those for Prime borrowers are \$26,300. These values vary monotonically across risk profiles, and suggest that though demand for credit card borrowing is higher for lower credit score borrowers, their access to credit card lending is severely restricted compared to borrowers in more favorable risk profiles.

The pattern is similar for mortgages. Only 14\% of Subprime borrowers have a mortgage, while 33\% of Prime borrowers do. Additionally, only 5\% of Subprime borrowers submit a mortgage inquiry in a quarter, and only 3\% of them show a mortgage origination in the same or subsequent quarter. By contrast, 6\% of Prime borrowers submit a mortgage inquiry in a quarter, and 9\% of them show an origination in the same or subsequent quarter. Consequently, average mortgage balances for Prime borrowers are \$66,420, while they are \$28,330 for Subprime borrowers. The variation in these statistics is monotone in the risk profile.

Figure \ref{fig:ccr_te} in Appendix \ref{app:costs} displays the time effects by risk profile. The time variation is interesting, as after the Great Recession, there was a tightening of standards for consumer loans, particularly mortgages. The figure shows that the gap in access to credit card and mortgage credit between borrowers at the bottom and at the top of the credit score distribution widened substantially from 2009 to 2016.

This analysis suggests that borrowers in non-Prime credit score risk profiles, who should have a more favorable ranking based on our model, have faced substantial restrictions in accessing credit card and mortgage loans. On the other hand, borrowers with Prime and Superprime scores, whose default probabilities are consistent with a worse risk rating based on our model, borrowed excessively, based on current industry standards, potentially exposing their lenders to unexpected losses.

\input{Tables/credit_risk_reg_l1_summary}

This exercise quantifies how, given current lending conditions, access to credit would change for consumers whose credit score risk profile differs from the one predicted by our model. Interest rates also vary significantly with credit scores, (\citeN{edelberg2006risk}), but since we do not have individual interest rate data, we are unable to document this variation in our sample. Changing the reference credit score and its performance properties would presumably lead to equilibrium changes in lending conditions. While we abstract from this possible effect, our analysis provides an assessment of how much being placed in the wrong risk profile can affect borrowers within the current system.

%% file: Tables/risk_def.tex
\begin{table}[htbp]\centering
\begin{threeparttable}
\footnotesize
\caption{Credit Score Risk Profiles}\label{tab:risk_def}
    \begin{tabular}{lccccc}
    \hline \hline
     & & & &  & \\[\dimexpr-\normalbaselineskip+2pt]
        & Deep Subprime & Subprime & Near Prime & Prime  & Super Prime  \\ 
             & & & &  & \\[\dimexpr-\normalbaselineskip+3pt] \hline
             & & & &  & \\[\dimexpr-\normalbaselineskip+3pt] 
        Credit score ranges & 300-499 & 500-600 & 601-660 & 661-780 & 781-850  \\ 
        Credit score percentiles  & 0-6.0 & 6.01-27.10 & 27.11-41.10 & 41.11-77.40 & 77.41-100  \\ 
        Percent of borrowers & 6.0 & 20.1 & 14.0 & 36.3 & 23.6 \\
        \hline \hline
    \end{tabular}
    \begin{tablenotes}
\footnotesize
\item Notes: Industry-defined credit score risk profiles. Definition by credit score range, credit score percentiles. All rates, fractions, and shares in percentage. Total \# of observations: 26,240,879. Time period 2006Q1-2016Q2. Source: Authors' calculations based on Experian Data.
\end{tablenotes}
\end{threeparttable}
\end{table}

%% file: Tables/risk_comp.tex
\begin{table}[!ht]
\begin{threeparttable}
\caption{Credit Score and Model Risk Profile Comparison}\label{tab:risk_comp}
    \centering
    \small
    \begin{tabular}{l|ccccc|c}
     
    & & & & &  & \\[\dimexpr-\normalbaselineskip+2pt]
     &\multicolumn{5}{c}{ Model Based Risk Profile} \\ 
            & & & & &  & \\[\dimexpr-\normalbaselineskip+2pt] \hline
            & & & & &  & \\[\dimexpr-\normalbaselineskip+2pt]
        Credit Score & Deep Subprime & Subprime & Near Prime & Prime  & Super Prime & Disagreement \\ 
        & & & & &  & \\[\dimexpr-\normalbaselineskip+2pt] \hline
Deep Subprime & 45.06 & 43.57 & 8.89 & 2.44 & 0.03 & 54.94 \\
Subprime & 14.67 & 52.71 & 22.46 & 10.07 & 0.08 & 47.29 \\
Near Prime & 1.66 & 38.05 & 30.35 & 28.08 & 1.86 & 69.65 \\
Prime & 0.17 & 4.43 & 12.83 & 66.50 & 16.07 & 33.50 \\
Super Prime & 0.02 & 0.05 & 0.15 & 25.71 & 74.08 & 25.92 \\
\hline \hline
    \end{tabular}
        
\begin{tablenotes}
\footnotesize
\item Notes: Rows correspond to industry-defined credit score risk profiles, defined in Table \ref{tab:risk_def}. Columns are fractions in each corresponding model-based risk profile. Column ``Disagreement" reports the fraction of borrowers in each credit score risk profile that would be placed in a different risk profile by our model. All values in percentage. Total \# of observations: 26,240,879. Time period 2006Q1-2016Q2. Source: Authors' calculations based on Experian Data.
\end{tablenotes}
\end{threeparttable}
\end{table}

%% file: Tables/risk_diff_new.tex
\begin{table}[htpb]\centering
\begin{threeparttable}
\caption{Default Risk Variation by Credit Score Profile}\label{tab:cs_comp}
\footnotesize
\begin{tabular}[c]{p{1.25in}p{0.75in}p{0.75in}p{1.25in}p{0.75in}p{0.75in}}
& & &  & & \\[\dimexpr-\normalbaselineskip+2pt]
 \multicolumn{1}{l}{Risk Profile} &   \multicolumn{2}{c}{Default Rate} &  \multicolumn{1}{l}{Risk Profile} &   \multicolumn{2}{c}{Default Rate}      \\
& & & & & \\[\dimexpr-\normalbaselineskip+2pt] \hline \hline 
& & &\\[\dimexpr-\normalbaselineskip+2pt]
 Credit Score &	 Realized 	&	 Predicted 	&	 Model Based 	 	&	 Realized 	&	 Predicted 		 	\\ 
\hline
& & &\\[\dimexpr-\normalbaselineskip+2pt]
\text{Deep Subprime} & 68.19 & 66.88 & \text{Deep Subprime} & 94.75 & 93.86  \\
& & & \text{Subprime} & 54.40 & 52.48 \\
& & & \text{Near Prime} & 18.18 & 17.75 \\
& & & \text{Prime} & 6.94 & 5.99 \\
& && \text{Super Prime} & 5.82 & 0.94 \\
\hline
& & & & &\\[\dimexpr-\normalbaselineskip+2pt]
\text{Subprime} & 43.68 & 43.10  & \text{Deep Subprime} & 92.48 & 92.46 \\
& & & \text{Subprime} & 48.25 & 47.59 \\
& & & \text{Near Prime} & 17.09 & 16.61 \\
& & & \text{Prime} & 8.27 & 7.20 \\
& & & \text{Super Prime} & 4.00 & 1.00 \\
\hline
& & &\\[\dimexpr-\normalbaselineskip+2pt]
\text{Near Prime} & 22.38   & 22.56 & \text{Deep Subprime} & 88.18 & 90.85 \\
& & & \text{Subprime} & 34.89 & 37.76 \\
 & & & \text{Near Prime} & 18.31 & 16.46 \\
& & & \text{Prime} & 7.36 & 5.97 \\
& & & \text{Super Prime} & 1.17 & 1.00 \\
\hline
& & & & &\\[\dimexpr-\normalbaselineskip+2pt]
\text{Prime} &  6.05 &6.62 & \text{Deep Subprime} & 90.87 & 91.53 \\
& & & \text{Subprime} & 25.88 & 32.72 \\
& & & \text{Near Prime} & 13.17 & 15.17 \\
& & & \text{Prime} & 4.38 & 4.40 \\
& & & \text{Super Prime} & 0.92 & 0.91 \\
\hline
& & & & &\\[\dimexpr-\normalbaselineskip+2pt]
\text{Super Prime} & 1.02 & 1.24 & \text{Deep Subprime} & 92.43 & 91.48 \\
& & & \text{Subprime} & 42.23 & 44.90 \\
& & & \text{Near Prime} & 10.15 & 14.44 \\
& & & \text{Prime} & 2.04 & 2.39 \\
& & & \text{Super Prime} & 0.61 & 0.76 \\
\hline \hline 
\end{tabular}
\begin{tablenotes}
\footnotesize
\item Notes: Realized and model predicted default rates by credit score and model-based risk profile. Industry-defined credit score risk profiles, defined in Table \ref{tab:risk_def}. All values in percentage. Total \# of observations: 26,240,879. Time period 2006Q1-2016Q2. Source: Authors' calculations based on Experian Data.
\end{tablenotes}
\end{threeparttable}
\end{table}

%% file: Tables/feature_attribution_differences.tex
\begin{table}[htbp]\centering
\begin{threeparttable}
\caption{Feature Attribution Differences}\label{tab:cs_vs_pp}
\small
\begin{tabular}[c]{p{1.5in}p{1.5in}p{1.25in}p{1.25in}}
 & & \multicolumn{2}{c}{\underline{Model}} \\ 
& & & \\
Feature Group & \# of Features &  Our Model & Credit Score  \\ \hline
& & & \\[\dimexpr-\normalbaselineskip+2pt]
Payment History & 21 & 0.33 & 0.35 \\
Amounts Owed & 43 & 0.49 & 0.3\\
Length of Credit History & 6 & 0.08 & 0.15 \\
Credit Mix & 5 & 0.05 & 0.1 \\
New Credit & 4 & 0.05 & 0.1 \\ \hline \hline 
\end{tabular}
\begin{tablenotes}
\footnotesize
\item Notes: SHAP values for five feature groups. For each prediction window, we compute the SHAP value for 100,000 randomly sampled observations and for each feature. We then calculate the sum of the absolute value for each feature, aggregate it across the feature groups, and report the results for the group. We normalized the results so that for each model the five groups sum up to 1. Feature groups are defined in Table \ref{tab:features} in Appendix \ref{app:model_details}. Credit score variation information from \url{https://www.myfico.com/credit-education/whats-in-your-credit-score}. Source: Authors' calculations based on Experian data.
\end{tablenotes}
\end{threeparttable}
\end{table}

%% file: Tables/credit_risk_reg_l1_summary.tex
\begin{table}[htbp]\centering
\begin{threeparttable}
\small
\def\sym#1{\ifmmode^{#1}\else\(^{#1}\)\fi}
\caption{Access to Credit by Risk Profile \label{tab:crr_reg_summary}}
\begin{tabular*}{\hsize}{@{\hskip\tabcolsep\extracolsep\fill}l*{4}{c}}
\hline\hline
                    &\multicolumn{1}{c}{(1)}&\multicolumn{1}{c}{(2)}&\multicolumn{1}{c}{(3)}&\multicolumn{1}{c}{(4)}\\
                    &\multicolumn{1}{c}{ Fraction with}&\multicolumn{1}{c}{Credit Limit / Balances}&\multicolumn{1}{c}{Inquiries}&\multicolumn{1}{c}{Originations if Inquired}\\
\hline
& & & & \\[\dimexpr-\normalbaselineskip+2pt]
\multicolumn{4}{l}{\underline{Panel A: Credit Cards}}\\
& & & & \\[\dimexpr-\normalbaselineskip+2pt]
Superprime & 96.13 & 39.43 & 5.90  & 35.15    \\
& & & & \\[\dimexpr-\normalbaselineskip+2pt]
Prime & 79.58 & 26.30 & 8.04 &  33.88   \\
& & & & \\[\dimexpr-\normalbaselineskip+2pt]
Near Prime & 66.88 & 19.76 & 12.37  & 29.87     \\
& & & & \\[\dimexpr-\normalbaselineskip+2pt]
Subprime & 38.15 & 7.73 & 11.40  & 16.70     \\
& & & & \\[\dimexpr-\normalbaselineskip+2pt]
Deep Subprime & 25.00 & 5.99 & 10.93  & 5.81       \\
\hline
& & & & \\[\dimexpr-\normalbaselineskip+2pt]
& & & & \\[\dimexpr-\normalbaselineskip+2pt]
\multicolumn{4}{l}{\underline{Panel B: Mortgages}}\\
& & & & \\[\dimexpr-\normalbaselineskip+2pt]
Superprime & 37.86 & 61.18 & 4.58  & 12.24    \\
& & & & \\[\dimexpr-\normalbaselineskip+2pt]
Prime & 32.93 & 66.42 & 5.99  & 9.08      \\
& & & & \\[\dimexpr-\normalbaselineskip+2pt]
Near Prime & 25.92 & 50.48 & 6.83  & 6.10     \\
& & & & \\[\dimexpr-\normalbaselineskip+2pt]
Subprime & 13.64 & 28.33 & 4.82  & 3.06     \\
& & & & \\[\dimexpr-\normalbaselineskip+2pt]
Deep Subprime & 11.14 & 25.55 & 4.15  & 1.77      \\
\hline
& & & & \\[\dimexpr-\normalbaselineskip+2pt]
\end{tabular*}
\begin{tablenotes}
\small
\item Notes: This table reports age-adjusted estimated coefficients for the average credit risk profile effects for credit cards (Panel A) and mortgages (Panel B). Deep Subprime, Subprime, Near Prime, Prime and Superprime are indicator variables for the consumer belonging to that category at quarter $t-1$ while outcomes are measured at quarter $t$. Column (2) is credit limits for credit cards and first mortgage balances for mortgages measured in thousands of USD. All other values in percentage. Column (4) restricts the sample to individuals who inquired in quarters $t-1,t$ and measures originations at quarter $t$. Full specification and regression results are reported in Appendix \ref{app:costs}.
\end{tablenotes}
\end{threeparttable}
\end{table}

%% file: Sections/vulnerable.tex
\section{Relation between Performance and Equity}\label{sec:vulnerable}

We have shown that ranking consumers by predicted default risk based on our model delivers significant performance improvements, particularly for borrowers at the bottom of the credit score distribution. However, there is growing concern about the potential negative impact of improved scoring algorithms for marginalized populations, such as young, minority, and low-income consumers. \citeN{fuster2018predictably} examine the racial implications of mortgage approvals based on a random forest model of default risk compared to a lower performing model based on logistic regression. They find that though each group gains from improved predictive accuracy, Black and White Hispanic borrowers lose standing relative to White and Asian borrowers with the more statistically advanced approach. They argue that the increased flexibility of the higher-performing model leads to an unequal distribution of the gains associated with improved performance. Additionally, \citeN{blattner2021costly} argue that sophisticated scoring models based on machine learning are not helpful at improving performance for marginalized borrowers because the main factor leading to worse performance for those borrowers is compositional differences in the attribute distribution, referred to as data bias.

We examine how using our model  to rank individuals by default risk affects the standing of different demographic groups traditionally marginalized in consumer credit markets in comparison to the credit score, and how the performance gains vary across these groups. We consider three criteria to identify marginalized borrowers: income, age and minority status.
Our credit report data contains limited demographic information, but does include date of birth and estimated annual income. This information allows us to construct indicator variables for being under 30 years old (Young) and for being in the bottom quintile of the income distribution (Income$_{p20}$). Our data does not include details on race due to its codification as a protected category in U.S. law. To address this, we construct a proxy for race derived from the 2010 Census data, where we combine Black and Hispanic populations into a single Minority indicator variable.\footnote{We construct our minority indicator following the approach proposed by the Consumer Financial Protection Bureau and described in \url{https://www.consumerfinance.gov/data-research/research-reports/using-publicly-available-information-to-proxy-for-unidentified-race-and-ethnicity/}. }
We also match our Experian data with  the Home Mortgage Disclosure Act (HMDA) that includes direct information on the race of mortgage borrowers and applicants. This restricts the sample to consumers with a first mortgage, but allows us to observe race directly. The details of our matching procedure are described in the Online Appendix.

\subsection{Variation in Standing}
To compare the standing of consumers in our model to the credit score, we rely on the percentile ranking introduced in Section \ref{sec:credit_score} and calculate the ranking differences between our model and the credit scores. A positive difference suggests that our model considers a borrower less risky than the credit score would indicate. This allows us to understand whether using our model's ranking generates disparities among different demographic groups, providing insights into the equity implications of improved performance.

Our analysis is based on regressions of the form:
\begin{equation}\label{reg:spec}
Y_{ist} = \alpha + \beta_1 x_{ist} + \beta_2 z_{ist} + \beta_3 z_{ist} \times x_{ist} +\lambda_t + \delta_s + \delta_{st} + \epsilon_{ist} 
\end{equation}
where $i$ corresponds to an individual, $s$ corresponds to the state, and $t$ corresponds to time in quarters. The main regressor is $x_{ist}$, an indicator for marginalized group membership. We include state-level controls as various regulations affecting the implication of default vary at the state level (\citeN{Albanesi_Nosal}). In selected specifications, we also control for realized default by including an indicator variable for whether an individual $i$ is ever 90 days or more past due in the subsequent 8 quarters at quarter $t$, the variable $z_{ist}$. Adding the default indicator as an explanatory variable allows us to assess how the differences in standing of consumers between our model and the credit score vary across consumers who default and those who do not. In most specifications, we also include an interaction term between $x_{ist}$ and $z_{ist}$, which allows us to evaluate whether rankings across the credit score and our model vary systematically for defaulting and non-defaulting consumers by demographic group. The variable $\lambda$ controls for any common time-varying factors, while $\delta$ controls for any time-invariant differences and any state-level trends, such as variations in economic conditions that might impact default probabilities. 

We conduct our analysis on the overall sample, known as the ``Experian Sample," using the race proxy and then replicate it with the matched data from both Experian and HMDA, termed the ``HMDA Match Sample." Descriptive statistics for these two samples by the demographic group are reported in Table \ref{tab:desc_vulnerable}. In each demographic group, borrowers in the HMDA Match sample have lower default rates, higher credit scores and higher household income. This suggests that they are positively selected relative to the overall Experian sample, as we would have expected, since appearing in the HMDA data requires that consumers have at least applied for a mortgage.
The descriptive statistics also reveal that younger, low income and minority borrowers tend to have higher default rates, lower total balances and overdue debt balances, and lower household incomes. They are also ranked lower both by the credit score and by our model. These differences appear both the Experian and in the HMDA Match samples.

Table \ref{tab:vulnerable_populations} reports our baseline regression results for both the Experian (Panel A) and HMDA Match (Panel B) samples. Column (1), where the only regressor in addition to the quarter$\times$state effects is a constant, confirms that for the overall sample there are no differences in consumer rankings on average, which is a consequence of our construction of the percentile rankings. Column (2) adds the future default indicator, which displays a negative and significant coefficient, with a value equivalent to approximately to 29 credit score points.\footnote{ One percentile in ranking is equivalent to approximately 5.5 credit score points.} The negative sign of this coefficient suggests that our model systematically ranks consumer who default lower than the credit score, consistent with its higher performance. Column (3) adds the low income indicator that displays a positive and significant coefficient. Its value implies that consumers in the lowest quantile of the income distribution would gain on average about 5 percentiles in the ranking with our model, equivalent to approximately 25 credit score points. Column (4) adds an interaction term between the low income and the future default indicators. The estimates for this specification suggest that low income consumers who do not default would gain 6 percentiles when ranked by our model, equivalent to approximately 33 credit score points, while those who default would lose 7 percentiles in the ranking, equivalent to approximately 39 credit score points. The estimates suggest that the improvement in standing of low income borrowers when ranked by our model is obtained by the model's better ability to separate those with high default risk from those with low default risk within that group. 

\input{Tables/vulnerable}

The findings are similar for young borrowers. The estimates in column (5) suggest our model would place young borrowers 2 percentiles higher in the ranking compared to the credit score, equivalent to approximately 11 credit score points. Column (6) adds the interaction with the default indicator, implying that young borrowers who do not default would gain 2 percentiles in rankings with our model, equivalent to 11 credit score points, but those who do default would lose 4.5 percentiles or 25 credit score points. Finally, column (7) deploys the Minority indicator. Overall, minority consumers would gain 1 percentile in ranking with our model. Column (8) introduces the interaction between minority and future default. Minority borrowers who do not default would gain 2 percentiles in the rankings with our model, equivalent to 11 credit score points, whereas those who default would lose 6 percentiles in the ranking, equivalent to 33 credit score points.

Taken together, the results for the Experian sample suggest that marginalized consumers, because of low income, youth or minority status, would on average gain in standing based on the ranking implied by our model compared to the credit score. The gain in ranking is driven by our model's superior ability to separate consumers by default risk in each demographic group. Findings are similar for the HMDA matched sample, as shown in Panel B. Consumers in this sample are positively selected by credit score and, consistent with our analysis in Section \ref{sec:credit_score}, our model tends to give them a lower ranking, as shown by the negative value of the constant term. However, low income and young borrowers in this sample gain on average in ranking, and these gains are concentrated among those who do not default, whereas those who default decline in ranking. The magnitude of the estimated coefficients is similar to the Experian sample. Minorities lose approximately 1 percentile in ranking in the HMDA sample, approximately 5 credit score points, even those who do not default, whereas those who default lose 3 percentiles in ranking equivalent to approximately 16 credit score points.

 Appendix \ref{app:tradeoff} reports the estimates for Prime and non-Prime borrowers.\footnote{Non-Prime consumers include those with a Deep Subprime, Subprime and Near Prime credit score, while Prime consumers include those with a Prime and Super Prime credit score.} Tables \ref{tab:vulnerable_populations_nonprime} shows that non-Prime consumers on average are ranked higher by our model by approximately 3 percentiles or 17 credit score points, with estimates by defaults status and demographic group very similar to the overall population. 
 Prime borrowers, as shown in Table \ref{tab:vulnerable_populations_prime}, are on average ranked 2 percentiles lower by our model, and low income, young and minority borrowers in this group do not gain in ranking compared to the credit score on average. However, both in the aggregate and for each demographic group, the estimates confirm that are model drops the ranking of those who will default, relative to the credit score, by approximately 7 percentiles.
 
 The results for non-Prime and Prime borrowers shed light on the discrepancy of the change in ranking for minorities in the Experian and HMDA Match sample. The latter, including only consumers with a mortgage, is a positively selected sample for conventional credit scoring models, as having a mortgage boosts the credit mix and typically improves the credit score. Credit mix is a less important factor for our model, as discussed in Section \ref{sec:credit_score}, which then tends to rank these consumers lower than the credit score. The positive selection associated with having a mortgage is likely stronger for minority consumers due to the history of redlining and housing discrimination in the United States. 

\subsection{Variation in Performance}
We now discuss performance by demographic group. Table \ref{tab:perf-vulnerables} reports AUC scores for the credit score and our model in the aggregate and for each of the marginalized groups we consider. For both the credit score and our model, statistical performance is lower for the marginalized groups compared to the rest of the population, though the gap in AUC scores between marginalized and non-marginalized groups is much larger for the credit score than for our model. For the overall sample, AUC is 5 percentage points higher for our model in comparison with the credit score, while the increase is much larger for the marginalized groups. For consumers in the bottom quintile of the income distribution, the AUC score rises to 0.86 for our model from 0.74 for the credit score, for consumers younger than 30 it rises to 0.88 for our model from 0.82 for the credit score, and for minority consumers it rises to 0.88 for our model from 0.80 for the credit score. The other consumers also experience an improvement in AUC score with our model compared to the credit score, though this is smaller, typically 4-5 percentage points, compared to 6-12 percentage points for the marginalized groups.

\input{Tables/perf_dem}

We now examine what drives this difference in performance. Table \ref{tab:SHAP-vulnerables} reports feature attribution statistics for each demographic group, based on SHAP values, showing virtually no difference across groups. This suggests that the conditional expectation function for default is similar across groups. Any differences in predictions are likely due to the distribution of attributes, consistent with \citeN{blattner2021costly}.


\input{Tables/feat_dem.tex}

\begin{figure}[!h]
	\centering
		
		\includegraphics[scale=0.5]{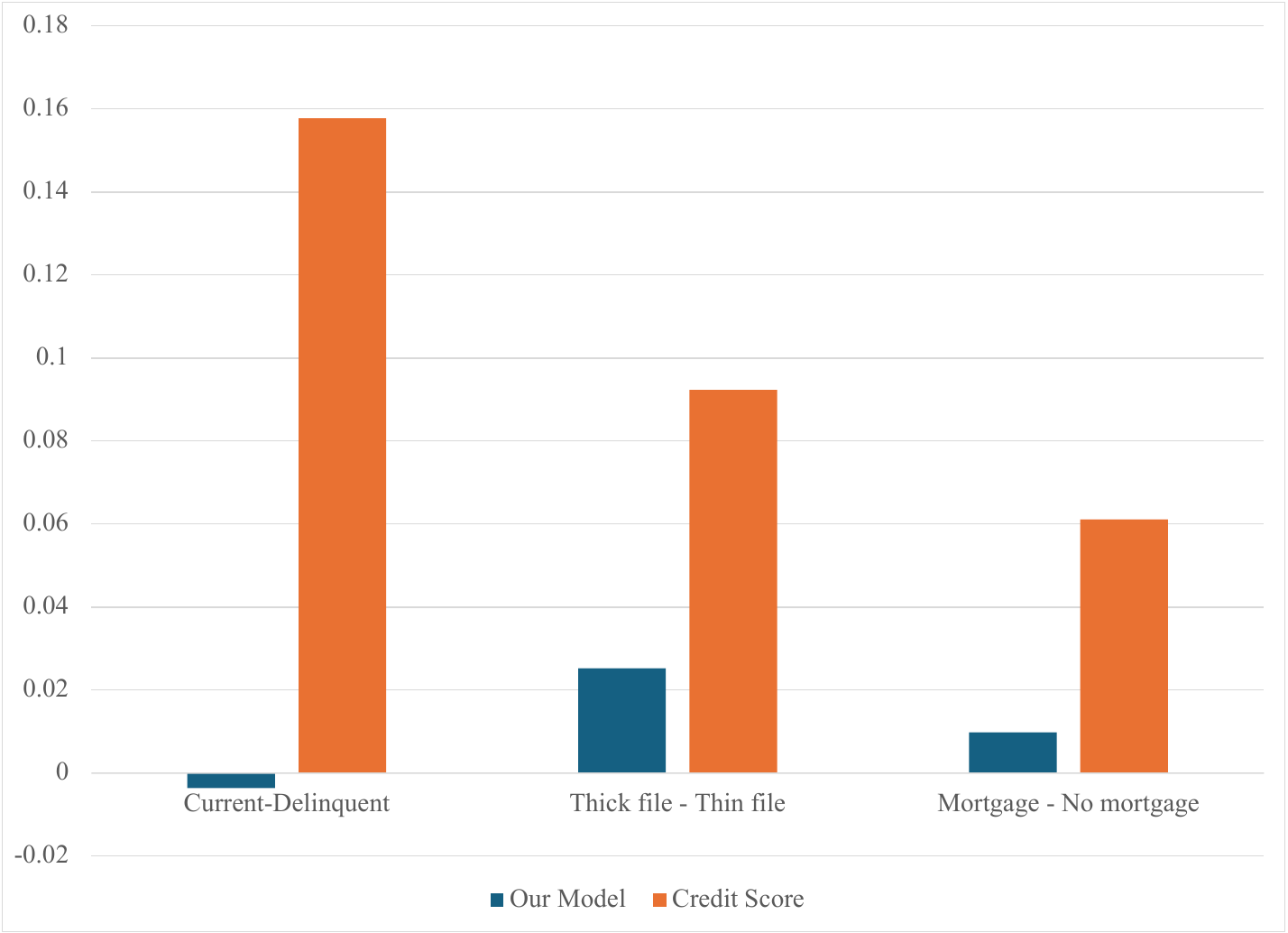}

\caption{Performance by Feature Composition}\label{fig:AUC_gap}
		\begin{flushleft}
		 \footnotesize{Notes: AUC scores for the credit score and our model's by feature composition. Averages 2006Q1-2016Q2. Source: Authors' calculations based on Experian data.}
		\end{flushleft}
\end{figure}

To examine the performance impact of the distribution of attributes across demographic groups, we calculate model and credit score AUC scores stratifying by attribute composition. The analysis is based on the notion that credit score performance is worse for consumers with delinquencies on their record, for those with thin files and for those with a limited credit mix, following \citeN{blattner2021costly}. To capture this, we calculate AUC scores for our model and the credit score separately for consumers who are current on all accounts and those who have any delinquencies, for those with thick files and those with thin files\footnote{Thin file consumers are those with a credit history of less than 10 years or less than 3 types of products on the credit report, among credit cards, bank cards, auto loans, HELOCs, first mortgages and installment loans. } and for those with a mortgage and those without a mortgage, where we interpret having a mortgage as another indicator of a broader credit mix, for each demographic group. Figure \ref{fig:AUC_gap} reports the difference in AUC scores across each feature composition category: Current-Delinquent, Thick files-Thin files, Mortgage - No mortgage. A positive gap suggests that performance is better for the favored feature composition category, such as current vs delinquent. The figure clearly illustrates that, for the credit score, performance as measured by the AUC score is significantly better for the favored attribute composition, while the differences are minimal for our model.

Table \ref{tab:feature_dis} reports the share of consumers in each feature composition category by demographic group. Low income, young and minority consumers are over-represented in the Delinquent, Thin File, No Mortgage categories, compared to older, higher income and non-minority consumers, with sizable differences in the share in each unfavorable feature composition group. The variation in AUC scores for our model and the credit score across the feature composition groups for each demographic group are reported in Table \ref{tab:AUC_feature} in Appendix \ref{app:tradeoff}. The gap in AUC scores across the favored and unfavored feature composition categories are similar across demographic groups for our model, and substantially smaller than for the credit score. 

\input{Tables/feat_comp.tex}

To evaluate how feature composition affects model performance for marginalized consumer groups, we conduct a straightforward counterfactual analysis. In this analysis, we calculate the AUC score for each marginalized group by assigning to it the same feature composition as the non-marginalized group. We then compare the difference between the actual and counterfactual AUC scores for both our model and the credit score. If the counterfactual AUC score is higher than the actual, it indicates that feature composition plays a role in reducing statistical performance. 
The results are presented in Table \ref{tab:AUC_CF} and show that the gap between the actual and counterfactual credit score are substantially smaller for our model compared to the credit score, in most cases by an order of magnitude. 
This suggests that for the credit score, feature composition is a significant factor lowering performance for  marginalized groups whereas this is not the case for our model.

\input{Tables/auc_cf.tex}

%
%
%

%
%
%
%
%



This analysis establishes that a credit scoring algorithm with superior statistical performance based on machine learning can yield greater performance improvements for groups traditionally marginalized on consumer credit markets, contrary to the findings in \citeN{blattner2021costly}. The performance improvements for marginalized group are driven by the model's improved predictive ability with low quality data, stemming from default status, thin files and low credit mix. The model's ability to minimize data bias raises performance for marginalized groups more than it does for the rest of the population and results in substantial improvements in ranking for these groups compared to the credit score. While these findings are promising, it is important to note that incorporating features more informative for young, low-income, or minority borrowers could potentially improve the AUC score for these marginalized groups even further, as suggested by \citeN{blattner2021costly}.

%% file: Tables/vulnerable.tex
\begin{table}[htbp]
\begin{center}
\begin{threeparttable}
\scriptsize
\def\sym#1{\ifmmode^{#1}\else\(^{#1}\)\fi}
\caption{Vulnerable Populations \label{tab:vulnerable_populations}}
\begin{tabular*}{\hsize}{@{\hskip\tabcolsep\extracolsep\fill}p{1.1in}*{8}{p{0.5in}}}
\hline\hline
& & & & & & & &\\[\dimexpr-\normalbaselineskip+2pt]
                    &\multicolumn{1}{c}{(1)}&\multicolumn{1}{c}{(2)}&\multicolumn{1}{c}{(3)}&\multicolumn{1}{c}{(4)}&\multicolumn{1}{c}{(5)}&\multicolumn{1}{c}{(6)}&\multicolumn{1}{c}{(7)}&\multicolumn{1}{c}{(8)}\\
\multicolumn{9}{l}{DV: Our Model Ranking - Credit Score Ranking}\\
\hline
& & & & & & & &\\[\dimexpr-\normalbaselineskip+2pt]
 \multicolumn{9}{l}{\underline{Panel A: Experian Sample}} \\ 
& & & & & & & &\\[\dimexpr-\normalbaselineskip+2pt]
Default  &                     &     -5.164\sym{***}&                     &     -6.392\sym{***}&                     &     -5.585\sym{***}&                     &     -5.111\sym{***}\\
                    &                     &    (0.032)         &                     &    (0.038)         &                     &    (0.036)         &                     &    (0.028)         \\
& & & & & & & & \\[\dimexpr-\normalbaselineskip+2pt]
Income$_{p20}$  &                     &                     &      4.485\sym{***}&      6.067\sym{***}&                     &                     &                     &                     \\
                    &                     &                     &    (0.069)         &    (0.077)         &                     &                     &                     &                     \\
& & & & & & & & \\[\dimexpr-\normalbaselineskip+2pt]
Income$_{p20}$ $\times$ Default &                     &                     &                     &     -0.438\sym{***}&                     &                     &                     &                     \\
                    &                     &                     &                     &    (0.048)         &                     &                     &                     &                     \\
& & & & & & & & \\[\dimexpr-\normalbaselineskip+2pt]
Young           &                     &                     &                     &                     &      2.296\sym{***}&      2.387\sym{***}&                     &                     \\
                    &                     &                     &                     &                     &    (0.070)         &    (0.074)         &                     &                     \\
& & & & & & & & \\[\dimexpr-\normalbaselineskip+2pt]
Young $\times$ Default &                     &                     &                     &                     &                     &      0.983\sym{***}&                     &                     \\
                    &                     &                     &                     &                     &                     &    (0.045)         &                     &                     \\
& & & & & & & & \\[\dimexpr-\normalbaselineskip+2pt]
Minority          &                     &                     &                     &                     &                     &                     &      0.998\sym{***}&      1.861\sym{***}\\
                    &                     &                     &                     &                     &                     &                     &    (0.035)         &    (0.041)         \\
& & & & & & & & \\[\dimexpr-\normalbaselineskip+2pt]
Minority $\times$ Default &                     &                     &                     &                     &                     &                     &                     &     -0.902\sym{***}\\
                    &                     &                     &                     &                     &                     &                     &                     &    (0.030)         \\
& & & & & & & & \\[\dimexpr-\normalbaselineskip+2pt]
Constant            &      0.001\sym{***}&      0.951\sym{***}&     -0.857\sym{***}&      0.047\sym{***}&     -0.498\sym{***}&      0.461\sym{***}&     -0.184\sym{***}&      0.643\sym{***}\\
                    &    (0.000)         &    (0.006)         &    (0.013)         &    (0.013)         &    (0.015)         &    (0.016)         &    (0.007)         &    (0.008)         \\ \hline  
& & & & & & & &\\[\dimexpr-\normalbaselineskip+2pt]
Observations        &26,240,879       &26,240,879       &26,240,879       &26,240,879       &26,240,879       &26,240,879       &26,240,879       &26,240,879       \\
$R^2$               &      0.0019         &      0.0181         &      0.0145         &      0.0391         &      0.0055         &      0.0228         &      0.0024         &      0.0196         \\
\hline 
& & & & & & & &\\[\dimexpr-\normalbaselineskip+2pt]
\multicolumn{9}{l}{\underline{Panel B: HMDA Match Sample}} \\
& & & & & & & & \\[\dimexpr-\normalbaselineskip+2pt]
Default     &                     &     -3.387\sym{***}&                     &     -3.790\sym{***}&                     &     -3.229\sym{***}&                     &     -3.532\sym{***}\\
                    &                     &    (0.052)         &                     &    (0.057)         &                     &    (0.059)         &                     &    (0.061)         \\
& & & & & & & & \\[\dimexpr-\normalbaselineskip+2pt]
Income$_{p20}$    &                     &                     &      2.979\sym{***}&      3.454\sym{***}&                     &                     &                     &                     \\
                    &                     &                     &    (0.096)         &    (0.105)         &                     &                     &                     &                     \\
& & & & & & & & \\[\dimexpr-\normalbaselineskip+2pt]
Income$_{p20}$ $\times$ Default &                     &                     &                     &     -0.131         &                     &                     &                     &                     \\
                    &                     &                     &                     &    (0.112)         &                     &                     &                     &                     \\
& & & & & & & & \\[\dimexpr-\normalbaselineskip+2pt]
Young                 &                     &                     &                     &                     &      2.430\sym{***}&      2.540\sym{***}&                     &                     \\
                    &                     &                     &                     &                     &    (0.084)         &    (0.088)         &                     &                     \\
& & & & & & & & \\[\dimexpr-\normalbaselineskip+2pt]
Young $\times$ Default &                     &                     &                     &                     &                     &     -0.709\sym{***}&                     &                     \\
                    &                     &                     &                     &                     &                     &    (0.099)         &                     &                     \\
& & & & & & & & \\[\dimexpr-\normalbaselineskip+2pt]
Minority         &                     &                     &                     &                     &                     &                     &     -1.113\sym{***}&     -0.973\sym{***}\\
                    &                     &                     &                     &                     &                     &                     &    (0.044)         &    (0.048)         \\
& & & & & & & & \\[\dimexpr-\normalbaselineskip+2pt]
Minority $\times$ Default &                     &                     &                     &                     &                     &                     &                     &      0.907\sym{***}\\
                    &                     &                     &                     &                     &                     &                     &                     &    (0.090)         \\
& & & & & & & & \\[\dimexpr-\normalbaselineskip+2pt]
Constant            &     -5.375\sym{***}&     -4.986\sym{***}&     -5.746\sym{***}&     -5.366\sym{***}&     -5.979\sym{***}&     -5.614\sym{***}&     -5.183\sym{***}&     -4.831\sym{***}\\
                    &    (0.000)         &    (0.006)         &    (0.012)         &    (0.014)         &    (0.021)         &    (0.024)         &    (0.008)         &    (0.010)         \\ \hline
& & & & & & & &\\[\dimexpr-\normalbaselineskip+2pt]
Observations        & 756,179      & 756,179      & 756,179      & 756,179      & 756,179      & 756,179      & 756,179      & 756,179      \\
$R^2$               &      0.0416         &      0.0487         &      0.0475         &      0.0564         &      0.0482         &      0.0556         &      0.0426         &      0.0494         \\
\hline\hline
\end{tabular*}
\begin{tablenotes}
\scriptsize
\item Notes: This table reports regressions of the form specified by Equation (\ref{reg:spec}). The dependent variable is the percentile ranking difference between our model and the credit score. The independent variable, "Default", is a binary indicator showing whether the individual defaulted within the next two years. The variable "Young" is a binary indicator for individuals under 30 years old. The variable "Income$_{p20}$" indicates whether the individual is in the lowest quintile of the estimated income distribution. In Panel A, the Minority indicator is based on ZIP code level Black and Hispanic population being over 50\%. In Panel B, the Minority indicator is 1 if the individual is Black or Hispanic. Panel A presents the full Experian sample, while Panel B provides the HMDA Match sample results. We control for State x Quarter fixed effects. Standard errors are clustered by state and quarter. \sym{*} \(p<0.10\), \sym{**} \(p<0.05\), \sym{***} \(p<0.01\).
\end{tablenotes}
\end{threeparttable}
\end{center}
\end{table}

%% file: Tables/perf_dem.tex
\begin{table}[!ht]
 
 \centering
 \begin{threeparttable}

 \caption{Performance by Demographic group} \label{tab:perf-vulnerables}
 \begin{tabular}{llllllll}
 \hline \hline
 AUC score & All & \multicolumn{2}{c}{Income } & \multicolumn{2}{c}{Age } & \multicolumn{2}{c}{Minority } \\ \hline
 ~ & ~ & $Income_{p20}$ & $Income_{\geq p20}$ & Age $< 30 $& Age $\geq 30$ & Yes & No \\ 
 Credit Score & 0.856 & 0.735 & 0.867 & 0.824 & 0.865 & 0.802 & 0.864 \\ 
 Our model & 0.906 & 0.855 & 0.908 & 0.881 & 0.913 & 0.881 & 0.91 \\ \hline \hline
 \end{tabular}
 \begin{tablenotes}
\footnotesize
\item Notes: AUC score for our model ad the credit score for each demographic group, averages for 2006Q1-2016Q2. Source: Authors' calculations based on Experian data.
\end{tablenotes}
\end{threeparttable}

\end{table}

%% file: Tables/feat_dem.tex
\begin{table}[!ht]
 \centering
 \begin{threeparttable}
 \caption{Feature Attribution by Demographic group} \label{tab:SHAP-vulnerables}

 \begin{tabular}{lllllll}
 \hline \hline

 Feature Group & \multicolumn{2}{c}{Income } & \multicolumn{2}{c}{Age } & \multicolumn{2}{c}{Minority } \\ \hline
 ~ & $Income_{p20}$ & $Income_{\geq p20}$ & Age $< 30 $& Age $\geq 30$ & Yes & No \\ 
 Payment History & 0.33 & 0.34 & 0.34 & 0.34 & 0.34 & 0.34 \\ 
 Amounts Owed & 0.49 & 0.49 & 0.49 & 0.49 & 0.49 & 0.49 \\ 
 Length of Credit History & 0.08 & 0.08 & 0.08 & 0.08 & 0.08 & 0.08 \\ 
 Credit Mix & 0.05 & 0.05 & 0.05 & 0.05 & 0.05 & 0.05 \\ 
 New Credit & 0.05 & 0.05 & 0.05 & 0.05 & 0.05 & 0.05 \\ \hline \hline
 \end{tabular}
 \begin{tablenotes}
\footnotesize
\item Notes: SHAP values for five feature groups. For each prediction window, we compute the SHAP value for 100,000 randomly sampled observations and for each feature. We then calculate the sum of the absolute value for each feature, aggregate it across the feature groups, and report the results for the group. We normalized the results so that the five groups sum up to 1. Feature groups are defined in Table \ref{tab:features} in Appendix \ref{app:model_details}. Time period 2006Q1-2016Q2. Source: Authors' calculations based on Experian data.
\end{tablenotes}
\end{threeparttable}
\end{table}

%% file: Tables/feat_comp.tex
\begin{table}[!ht]
 \centering
 \begin{threeparttable}
 \caption{Feature Composition by Demographic Group} \label{tab:feature_dis}
 \small
 \begin{tabular}{llcccccc}
 \hline \hline
 & \multicolumn{7}{c}{Demographic Group } \\ 
 & & \multicolumn{2}{c}{Income } & \multicolumn{2}{c}{Age } & \multicolumn{2}{c}{Minority } \\ \hline
 Feature Category & Status & $Income_{p20}$ & $Income_{\geq p20}$ & Age $< 30 $& Age $\geq 30$ & Yes & No \\ \\ 
 Default History & Current & 0.64 & 0.87 & 0.80 & 0.83 & 0.73 & 0.84 \\ 
 ~ & Delinquent & 0.36 & 0.13 & 0.20 & 0.17 & 0.27 & 0.16 \\ 
 Credit History & Thin file & 0.93 & 0.39 & 0.89 & 0.39 & 0.64 & 0.47 \\ 
 ~ & Thick file & 0.07 & 0.61 & 0.11 & 0.61 & 0.36 & 0.53 \\ 
 Mortgage Status & No Mortgage & 0.95 & 0.61 & 0.91 & 0.62 & 0.78 & 0.66 \\ 
 ~ & Mortgage & 0.05 & 0.39 & 0.09 & 0.38 & 0.22 & 0.34 \\ \hline \hline
 \end{tabular}
 \begin{tablenotes}
\footnotesize
\item Notes: Share of borrowers in each feature composition category by demographic group. Feature composition categories defined by Default History, Credit History, Mortgage Status and Credit History interacted with Mortgage Status, with shares within each category adding up to 1. Time period 2006Q1-2016Q2. Source: Authors' calculations based on Experian data.
\end{tablenotes}
\end{threeparttable}

\end{table}

%% file: Tables/auc_cf.tex
\begin{table}[!ht]
 \centering
 \begin{threeparttable}
 \caption{Contribution of Feature Composition to Model Performance} \label{tab:AUC_CF}

 \begin{tabular}{llccc} \hline \hline

 Group & & Default & Credit History & Mortgage Status \\ \hline

 Young & Our Model& -0.001 & 0.014 & 0.004 \\ 
 & Credit Score & 0.004 & 0.040 & 0.018 \\ 
 $I_{p20}$ & Our Model & -0.017 & -0.005 & 0.006 \\ 
 & Credit Score & 0.012 & 0.040 & 0.035 \\ 
 Minority & Our Model & -0.005 & 0.004 & 0.001 \\ 
 & Credit Score & 0.013 & 0.019 & 0.009 \\ \hline \hline

 \end{tabular}

 \begin{tablenotes}
\footnotesize
\item Notes: Gap between counterfactual and actual AUC score for marginalized consumers by group for our model and the credit score by feature component. Counterfactual AUC score applies feature distribution of non-marginalized group to marginalized group. Time period 2006Q1-2016Q2. Source: Authors' calculations based on Experian data.
\end{tablenotes}
\end{threeparttable}
\end{table}

%% file: Sections/conclusion.tex
\section{Conclusion}
Our analysis demonstrates that widely used credit scores exhibit a significant degree of misclassification, wherein a substantial proportion of consumers are assigned to risk categories that do not accurately reflect their true default risk. This phenomenon is particularly pronounced among individuals with lower credit scores.
Furthermore, our research indicates that the utilization of a machine learning-based scoring algorithm, characterized by superior predictive performance even with low data quality, would result in a more favorable credit risk assessment for low-income, young, and minority consumers. This suggests that the implementation of such an algorithm has the potential to ameliorate existing disparities in credit access, promoting a more equitable allocation of consumer credit.

%% file: Sections/appendix.tex
\appendix
\addcontentsline{toc}{section}{Appendices}
\section*{Appendix}

\section{Model Details}\label{app:model_details}

We use a Hybrid Deep Neural Network-Gradient Boosted Trees model for our analysis. This is an ensemble model, comprised of two components.
The first is based on deep learning, in the class used by \citeN{sirignano}. We restrict attention to feed-forward neural networks, composed of an input layer, which corresponds to the data, one or more interacting hidden layers that non-linearly transform the data, and an output layer that aggregates the hidden layers into a prediction. Layers of the networks consist of neurons, with each layer connected by synapses that transmit signals among neurons of subsequent layers. A neural network is, in essence, a sequence of nonlinear relationships. Each layer in the network takes the output from the previous layer and applies a linear transformation followed by an element-wise nonlinear transformation.

The second component of our model is Extreme Gradient Boosting, which builds on decision tree models. Tree-based models split the data several times based on certain cutoff values in the explanatory variables.\footnote{Splitting means that different subsets of the dataset are created, where each observation belongs to one subset.} Gradient Boosted Trees (GBT) are an ensemble learning method designed to correct the tendency of tree-based models to overfit training data. This is achieved by recursively combining the predictions of multiple, simpler trees. Although each individual shallow tree is a ``weak learner" with limited predictive power, the ensemble of these weak learners forms a strong model with improved stability over a single complex tree.

These two components — Deep Neural Network (DNN) and Gradient Boosted Trees (GBT) — are combined to improve predictive performance. Each model is trained independently, and their final predicted probabilities are then averaged. This approach, akin to the method proposed by \citeN{kvamme2018}, which combined a convolutional neural network with a random forest by averaging, ensures that our methodology benefits from the strengths of both models. Specifically, we achieve this by following two key steps:

\begin{enumerate}
 \item For each observation, run DNN and GBT separately and obtain predicted probabilities for each of the models;
 \item Compute a weighted average of the predicted probabilities, with the weights determined based on model performance, as explored in \citeN{albanesi_vamossy_NBER_2019}.
\end{enumerate}

Table \ref{tab:features} lists the features from the model we use as inputs from the credit report data. They include information on balances and credit limits for different types of consumer debt, severity and number of delinquencies, credit utilization by type of product, public record items such as bankruptcy filings, collection items, and length of the credit history. In order to be consistent with the restrictions of the Fair Credit Reporting Act of 1970 and the Equal Opportunity in Credit Access Act of 1984, we do not include information on age or zip code, and we do not include any income information, to be consistent with current credit scoring models. Table \ref{tab:features} lists the complete set of features used in our machine learning models.

\input{Tables/features}

\clearpage
\section{Comparison with Credit Scores}\label{app:score}

The credit score is a summary indicator used to predict the likelihood of a borrower defaulting, and it is widely used in the financial industry. For most unsecured debts, lenders typically check a prospective borrower's credit score at the time of application. They may also review a recent short sample of their credit history. For larger unsecured debts, as well as secured debts like mortgages and auto loans, lenders usually require some form of income verification. Nevertheless, the credit score often plays a crucial role in determining key terms of the borrowing contract, such as the interest rate, down payment, or credit limit. 

The most widely known credit score is the FICO score, a measure generated by the Fair Isaac Corporation, which has existed since 1989. Each of the three major credit reporting bureaus-- Equifax, Experian, and TransUnion-- also develops its proprietary credit scores. Credit scoring models are not public, though they are restricted by the law, mainly the Fair Credit Reporting Act of 1970 and the Consumer Credit Reporting Reform Act of 1996. These laws mandate that consumers are informed of the four main factors affecting their credit score.
According to available materials from FICO and the credit bureaus, payment history and outstanding debt together account for more than 65\% of the variation in credit scores. This is followed by the length of credit history, which explains 15\% of the variation, and by new accounts and types of credit used (10\%), along with new ``hard" inquiries —credit report inquiries initiated by prospective lenders.

\begin{figure}[!h]
\centering
\includegraphics[scale=0.5]{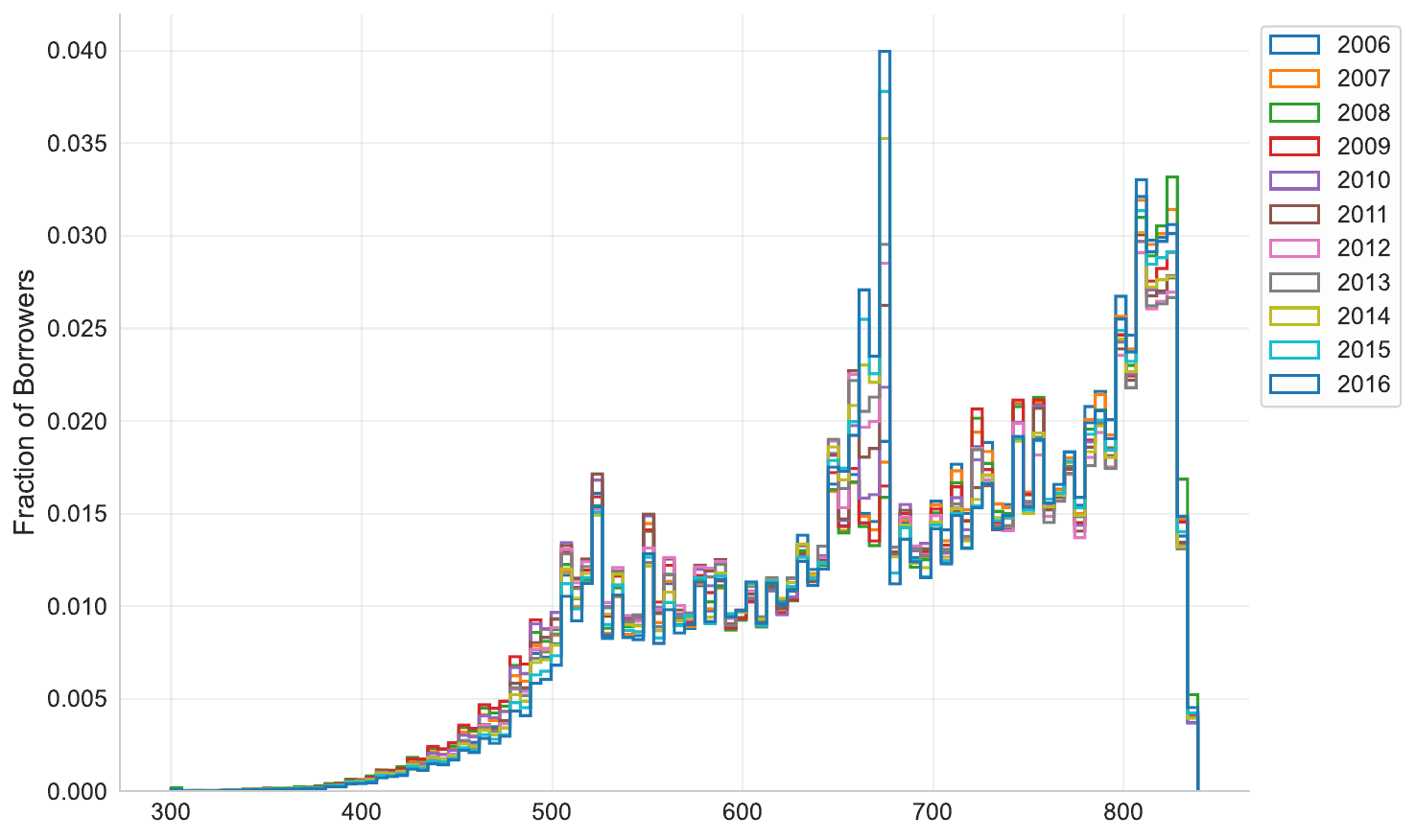}
\caption{Credit Score Histogram by Years}\label{fig:vs_hist}
		\begin{flushleft}
		 \footnotesize{Notes: Histogram of the credit score in our data by year for selected years. Source: Authors' calculations based on Experian Data.}
		\end{flushleft}
\end{figure}

U.S. law prohibits credit scoring models from considering a borrower's race, color, religion, national origin, sex and marital status, age, address, public assistance receipt, or any consumer right exercise under the Consumer Credit Protection Act. The credit score cannot be based on information not found in a borrower's credit report, such as salary, occupation, title, employer, date employed or employment history, or interest rates charged on particular accounts.  Additionally, items related to child or family support obligations and ``soft" inquiries—including ``consumer-initiated" inquiries (e.g., requests to view one's credit report), ``promotional inquiries" (e.g., pre-approved credit offers), or ``administrative inquiries" (e.g., lender reviews of open accounts) — are excluded from the scoring model. Requests marked as coming from employers are also not counted. In general, any information that is not proven to be predictive of future credit performance cannot be included in the credit score calculation.

\begin{figure}[!h]
	\centering

	\begin{subfigure}[b]{0.48\textwidth}
		\includegraphics[scale=0.3]{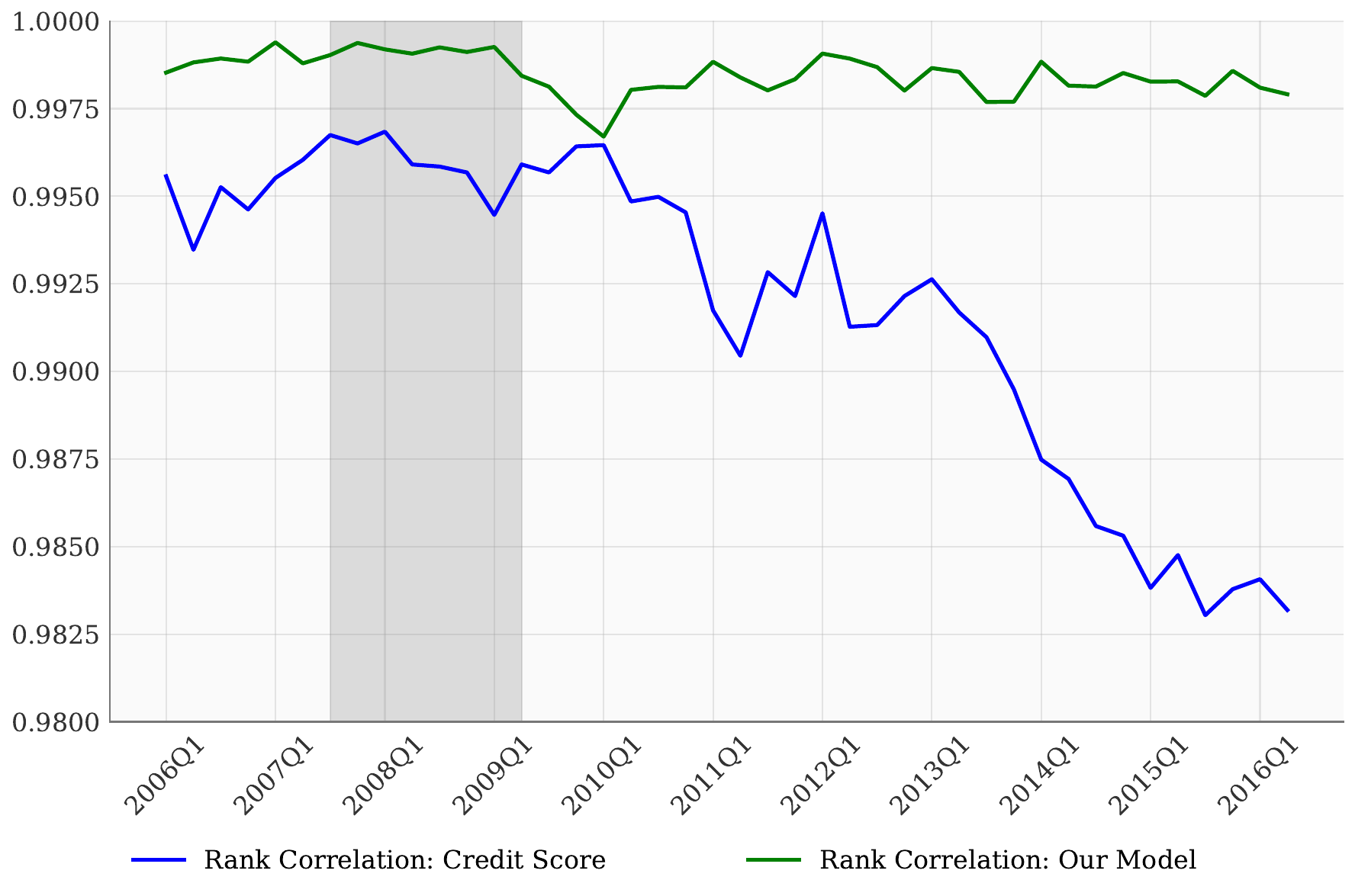} 
		\caption{Spearman Rank Correlation}
	\end{subfigure}
 	\begin{subfigure}[b]{0.48\textwidth}
		\includegraphics[scale=0.3]{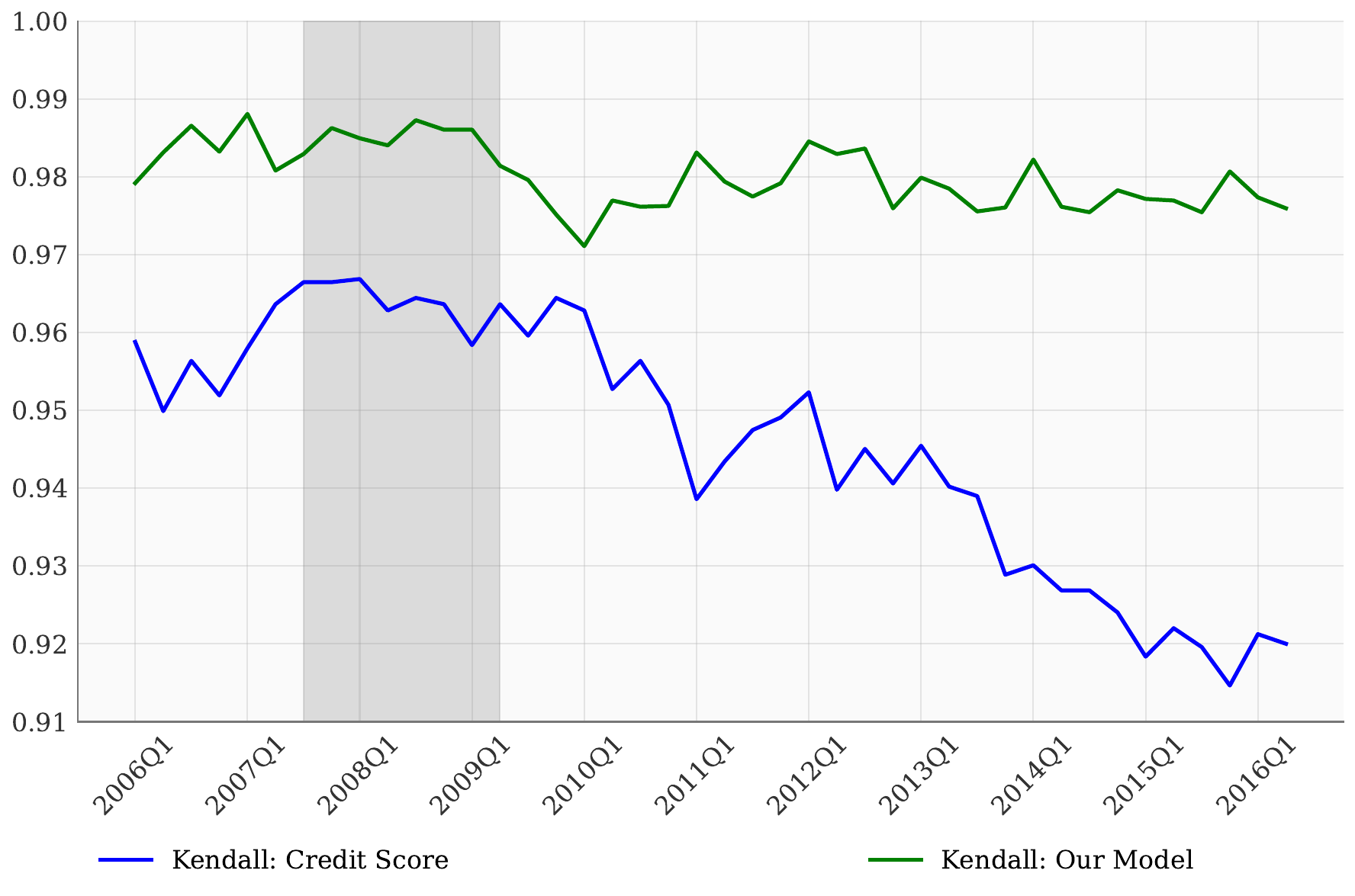} 
		\caption{Kendall Correlation}
	\end{subfigure}

\caption{Gini Coefficient and Rank Correlation with Realized Default Rate}\label{fig:rank_corr_comp}
		\begin{flushleft}
		 \footnotesize{Notes: Panel a): Spearman's rank correlation for the credit score and our model's by quarter. Panel b): Kendall correlation for the credit score and our model's by quarter. Source: Authors' calculations based on Experian data.}
		\end{flushleft}
\end{figure}

\clearpage

\section{Costs of Misclassification}\label{app:costs}
\input{Tables/misclassification_desc}

\clearpage
\subsection{Access to Credit}
\begin{sidewaysfigure}
 \begin{subfigure}{0.55\textwidth}
 \centering
 \includegraphics[width=\textwidth]{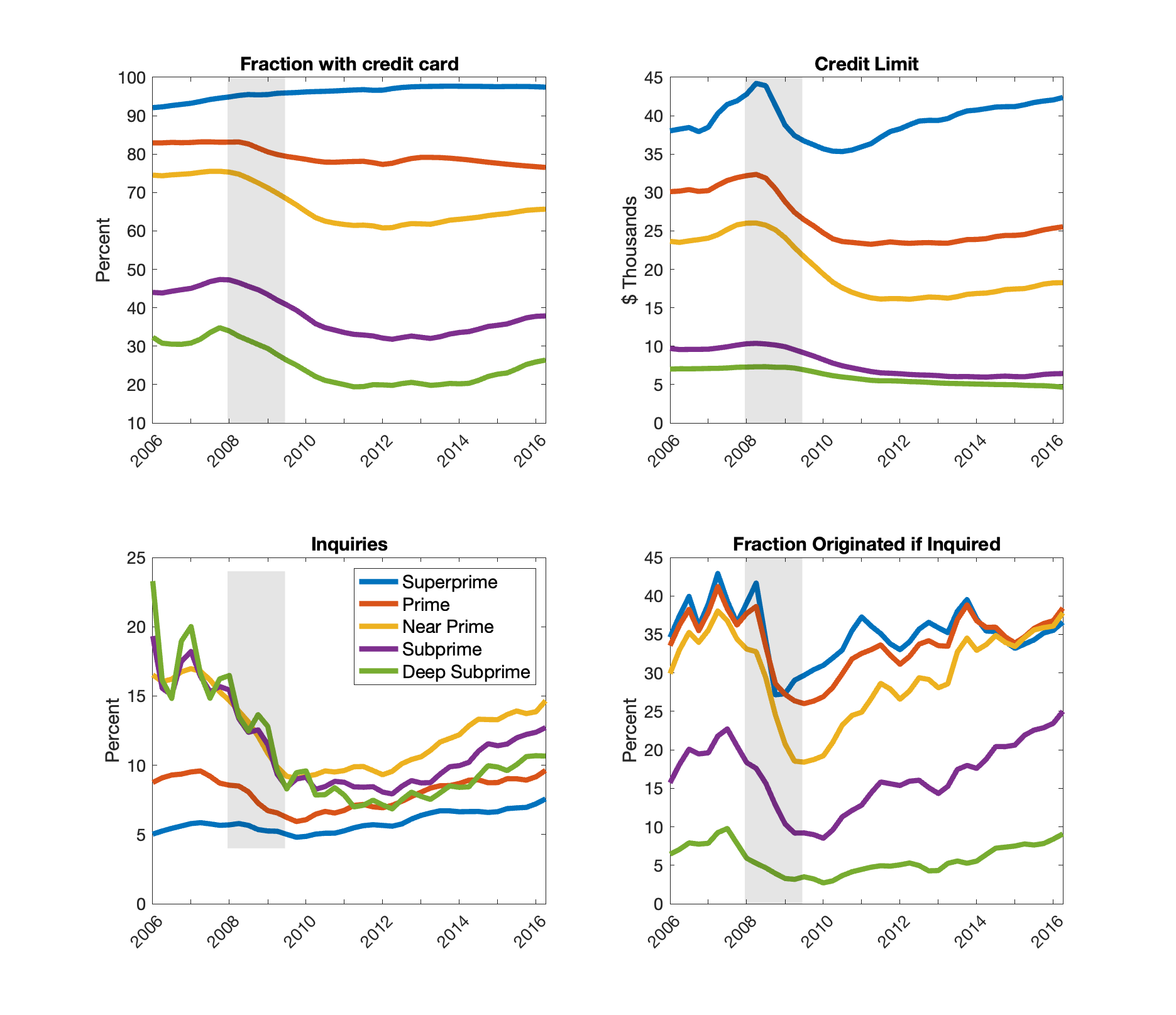} 
 \caption{Credit Cards}
 \label{fig:chart1}
 \end{subfigure}
 \begin{subfigure}{0.55\textwidth}
 \centering
 \includegraphics[width=\textwidth]{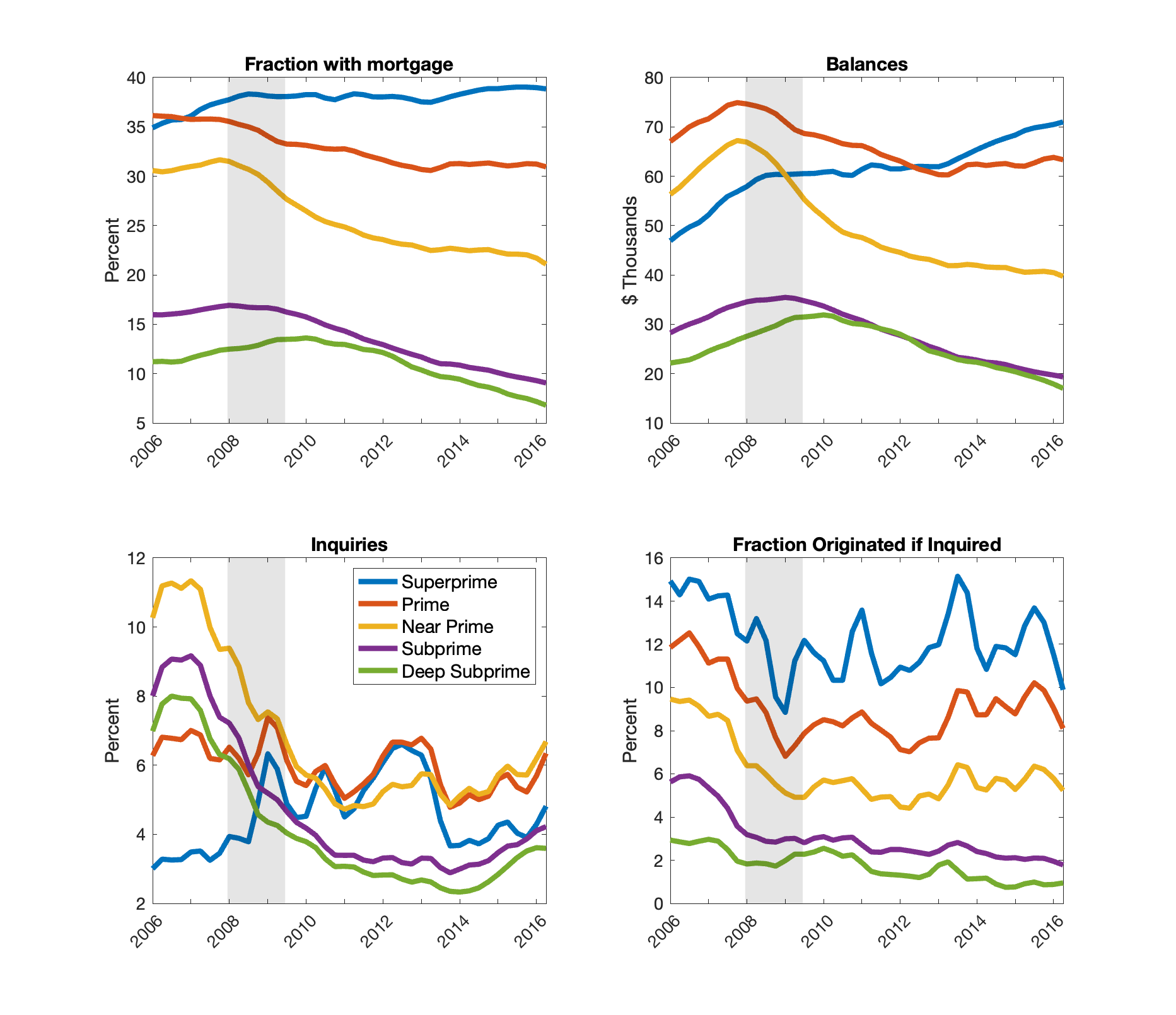} 
 \caption{Mortgages}
 \label{fig:chart2}
 \end{subfigure}
 \caption{Credit Access By Risk Profile Overtime} 
 \label{fig:ccr_te}
 
 \begin{flushleft}
 {\footnotesize Age-adjusted quarter effects interacted with credit risk profile from estimation of equation (\ref{reg:access_spec}) for credit cards and mortgages. Age adjustment obtained by assuming aggregate age average age distribution for each credit risk profile and using age bin effects reported in Table \ref{tab:crr_reg}. See text for more detail.}
 \end{flushleft}

 \end{sidewaysfigure}

The specification for the access to credit regressions in Section \ref{sec:value_added_consumers} is as follows:
\begin{equation}\label{reg:access_spec}
y_{it} = \alpha + \beta crp_{i,t-1} + \gamma a_{it} + \lambda_t \times crp_{i,t-1} + \epsilon_{it} 
\end{equation}
where $i$ corresponds to an individual and $t$ corresponds to time in quarters. The main regressor is $crp_{i,t-1}$, an indicator for credit score risk profile at $t-1$, varying from Deep Subprime to Superprime. The credit profile is measured at $t-1$ to avoid joint endogeneity in risk profile and credit outcomes (see \citeN{albanesi2022credit} for a discussion). The variable $a_{it}$ is an indicator for the consumer's age bin. We allow for five age bins, corresponding to the following age ranges in years: 18-29, 30-39, 40-49, 50-59, 60-69 and 70+. The variable $\lambda$ controls time-varying factors affecting access to credit over the sample period, and is interacted with the credit risk profile, to detect any risk profile specific variations in access to credit over the sample period. Additionally, we include zip code fixed effect to account for geographical variation in home values and consumer prices and other factors that might affect the level of credit card and mortgage balances.
We consider four possible outcomes: credit limits for credit cards and first mortgage balances for mortgages; inquiries and originations for credit cards and mortgages; and originations at quarter $t$ among consumers with inquiries at quarter $t-1,t$, to allow for a lag between the posting of an inquiry and the possible realization of an origination. The estimated are reported in Table \ref{tab:crr_reg}. The age adjusted estimated values for the time effects interacted with the credit risk profile are displayed in Figure \ref{fig:ccr_te}.

\input{Tables/credit_risk_reg.tex}

\clearpage

\section{Trade-off Between Performance and Equity}\label{app:tradeoff}

\begin{table}[!ht]
\centering
 \begin{threeparttable}
 \caption{AUC by Feature Composition} \label{tab:AUC_feature}
\small

 \begin{tabular}{llcccccc}
 \hline \hline
 \multicolumn{8}{c}{AUC Score: Our Model } \\ \hline

 Feature Category & Status & \multicolumn{2}{c}{Income } & \multicolumn{2}{c}{Age } & \multicolumn{2}{c}{Minority } \\ \hline

 & & $Income_{p20}$ & $Income_{\geq p20}$ & Age $< 30 $& Age $\geq 30$ & Yes & No \\ 

 Default History & Current & 0.81 & 0.86 & 0.78 & 0.85 & 0.81 & 0.85 \\ 
 ~ & Delinquent & 0.84 & 0.85 & 0.85 & 0.85 & 0.85 & 0.85 \\ 
 Credit History & Thin file & 0.88 & 0.90 & 0.85 & 0.90 & 0.87 & 0.90 \\ 
 ~ & Thick file & 0.90 & 0.92 & 0.84 & 0.91 & 0.90 & 0.92 \\ 
 Mortgage Status & No Mortgage & 0.88 & 0.91 & 0.85 & 0.91 & 0.88 & 0.91 \\ 
 ~ & Mortgage & 0.89 & 0.91 & 0.87 & 0.91 & 0.89 & 0.91 \\ 
 &&&&&&& \\ \hline \hline
 \multicolumn{8}{c}{AUC Score: Credit Score } \\ \hline

 Feature Category & Status & \multicolumn{2}{c}{Income } & \multicolumn{2}{c}{Age } & \multicolumn{2}{c}{Minority } \\ \hline

 & & $Income_{p20}$ & $Income_{\geq p20}$ & Age $< 30 $& Age $\geq 30$ & Yes & No \\ 
 
 Default History & Current & 0.76 & 0.81 & 0.66 & 0.80 & 0.74 & 0.81 \\ 
 ~ & Delinquent & 0.64 & 0.65 & 0.61 & 0.67 & 0.63 & 0.65 \\ 
 Credit History & Thin file & 0.82 & 0.81 & 0.74 & 0.83 & 0.77 & 0.82 \\ 
 ~ & Thick file & 0.89 & 0.91 & 0.81 & 0.90 & 0.88 & 0.91 \\ 
 Mortgage Status & No Mortgage & 0.82 & 0.85 & 0.73 & 0.85 & 0.79 & 0.85 \\ 
 ~ & Mortgage & 0.88 & 0.90 & 0.84 & 0.90 & 0.87 & 0.90 \\ \hline \hline
 \end{tabular}
 
 \ssk
 
 \begin{tablenotes}
\footnotesize
\item Notes: AUC score for marginalized consumers by group for our model and the credit score by feature component. Source: Authors' calculations based on Experian data.
\end{tablenotes}
\end{threeparttable}

\end{table}

\input{Tables/vulnerable_desc}

\input{Tables/vulnerable_nonprime}

\input{Tables/vulnerable_prime}

\clearpage
\input{Sections/hmda_match}

%% file: Tables/features.tex
\begin{table}[htbp]\centering
\scriptsize
\begin{threeparttable}
\caption{Model Inputs}\label{tab:features}
\begin{tabular}{p{2.5in}p{3.5in}}
\hline
& \\
Amt. past due on bankcard revolving and charge trades presently 30 days delinquent & Monthly pmt. on open non-def. student trades            \\
Amt. past due on credit card trades 90-180 DPD+     & Monthly pmt. on second mortgage trades                \\
Amt. past due on installment trades 90-180 DPD+     & Mo. since the most recent 30-180 DPD+ on credit card trades     \\
Amt. past due on joint mortgage type trades              & Mo. since the most recent 30-180 DPD+ on trades         \\
Amt. past due on revolving trades presently 30 days delinquent       & Mo. since the most recent 90+ days delinquency              \\
Amt. past due on revolving trades 90-180 DPD+       & Mo. since the most recent foreclosure proceeding started on first mortgage trades \\
Amt. past due on trades presently 30 DPD+          & Mo. since the most recently closed, transferred, or refinanced first mortgage trade \\
Amt. past due on trades presently 90-180 days delinquent         & Mo. since the most recently opened credit card trade            \\
Balance on bankcard revolving and charge trades 90-180 DPD+  & Mo. since the most recently opened first mortgage trade           \\
Balance on collections                       & Mo. since the most recently opened HELOC trade       \\
Balance on collections, placed with the collector in the last 24m      & Mo. since the oldest trade was opened                 \\
Balance on credit \& bankcards                   & Mortgage to total debt                       \\
Balance on HELOC trades               & Mortgage type inquiries made inthe last 3 m             \\
Balance on installment trades                    & No. of auto loan trades                     \\
Balance on installment trades presently 90-180 days delinquent         & No. of collections                        \\
Balance on open auto loan trades                   & No. of credit \& bankcards                    \\
Balance on revolving trades presently 90-180 days delinquent         & No. of installment trades                     \\
Balance on second mortgage trades                  & No. of occurrencies of 90 days delinquencies in the last 36m      \\
Balance on trades presently 30 days delinquent             & No. of occurrencies of 90 days delinquencies in the last 6m       \\
Balance on trades presently 60 days delinquent             & No. of open mortgage type trades                  \\
Balance on trades presently 90+ days delinquent or derogatory        & Open HELOC trades                 \\
Bankcard revolving and charge inquiries made in the last 3m        & Public record bankruptcies                     \\
Credit amt. on  HELOC             & Public record discharged bankruptcies                  \\
Credit amt. on open credit card trades                 & Public record dismissed bankruptcies                   \\
Credit amt. on open def. student trades              & Public records filed in the last 24m                \\
Credit amt. on open non-def. student trades            & Ratio of inquiries (no deduplication) to trades opened in the last 6m    \\
Credit amt. on open trades                     & Total debt balances                        \\
Credit amt. on revolving trades                  & Trades legally paid in full for less than the full balance           \\
Credit amt. paid down on open first mortgage trades            & Unsatisfied collections                      \\
Credit card utilization                      & Utilization ratio                        \\
Fraction of 30 days delinquent debt                  & Worst ever status on a credit card trade in the last 24m         \\
Fraction of 60 days delinquent debt                  & Worst ever status on a mortgage type trade in the last 24m         \\
Fraction of 90+ days delinquent debt                 & Worst ever status on a trade in the last 24m             \\
Heloc utilization                        & Worst present status on a credit card trade                \\
Inquiries made in the last 12m (no deduplication)          & Worst present status on a mortgage type trade              \\
Installment utilization                      & Worst present status on a trade                    \\
Monthly pmt. on credit card trades                  & Worst present status on a trade (excl. collections)            \\
Monthly pmt. on debt                      & Worst present status on an installment trade               \\
Monthly pmt. on joint mortgage type trades              & Worst present status on an open trade                  \\
Monthly pmt. on open first mortgage trades              &  \\ \hline  \hline 
\end{tabular}
\begin{tablenotes}
\scriptsize
\item Notes: Amt=Amount, pmt=Payment, dpd=Days past due, HELOC=Home equity line of credit, def.=Deferred, revolving=Revolving credit, Mo.=Months, m=Months, No.=Number, excl.=Excluding.
\end{tablenotes}
\end{threeparttable}
\end{table}

%% file: Tables/misclassification_desc.tex
\begin{table}[htbp]\centering
\begin{threeparttable}
\caption{Descriptive Statistics on Credit Access by Credit Score Risk Profile}\label{tab:misclass_desc}
\footnotesize
\begin{tabular}{lcccccc}
\hline\hline
& & & & & & \\[\dimexpr-\normalbaselineskip+2pt]
                    & \multicolumn{1}{c}{(1)} & \multicolumn{1}{c}{(2)} & \multicolumn{1}{c}{(3)} & \multicolumn{1}{c}{(4)} & \multicolumn{1}{c}{(5)} & \multicolumn{1}{c}{(6)} \\
                    & \multicolumn{1}{c}{Credit Limit} & \multicolumn{1}{c}{\% Inquired} & \multicolumn{1}{c}{ Origination} & \multicolumn{1}{c}{Origination} & \multicolumn{1}{c}{\% of Borrowers} & \multicolumn{1}{c}{Credit Limit} \\
                    & \multicolumn{1}{c}{/ Balances} & \multicolumn{1}{c}{} & \multicolumn{1}{c}{} & \multicolumn{1}{c}{if Inquired} & \multicolumn{1}{c}{with} & \multicolumn{1}{c}{or Balances /} \\
                   & & & & & & \multicolumn{1}{c}{Income (\%)} \\
\hline
& & & & & & \\[\dimexpr-\normalbaselineskip+2pt]
\multicolumn{7}{l}{\underline{Panel A: Credit Cards}} \\
& & & & & & \\[\dimexpr-\normalbaselineskip+2pt]
Deep Subprime & 7,031.5 & 11.4 & 6.2 & 2.4 & 1.36 & 12.0 \\
Subprime & 13,170.4 & 11.6 & 17.3 & 6.1 & 7.36 & 20.1 \\
Near Prime & 27,685.4 & 12.6 & 31.0 & 13.2 & 9.30 & 36.0 \\
Prime & 32,923.3 & 8.2 & 34.5 & 13.6 & 28.94 & 35.8 \\
Superprime & 44,710.7 & 5.0 & 34.9 & 14.0 & 23.01 & 39.1 \\
& & & & & & \\[\dimexpr-\normalbaselineskip+2pt]
\multicolumn{7}{l}{\underline{Panel B: Mortgages}} \\
& & & & & & \\[\dimexpr-\normalbaselineskip+2pt]
Deep Subprime & 171,036.42 & 4.7 & 0.5 & 2.3 & 0.56 & 268.02 \\
Subprime & 172,753.09 & 5.0 & 0.8 & 3.9 & 2.62 & 238.81 \\
Near Prime & 188,108.00 & 6.9 & 1.6 & 7.6 & 3.60 & 224.40 \\
Prime & 208,991.91 & 6.0 & 1.9 & 10.6 & 11.79 & 195.83 \\
Superprime & 170,236.19 & 4.1 & 2.1 & 13.4 & 9.30 & 123.35 \\
\hline
\end{tabular}
\begin{tablenotes}
\footnotesize
\item Notes: The table presents descriptive statistics categorized by credit score risk profiles from 2006Q1 to 2016Q2. The credit score risk profiles are defined in Table \ref{tab:risk_def}. The sample consists of 26,147,712 borrowers. The credit risk profile is determined based on the previous quarter. \% Originated represents the fraction of borrowers within each credit risk profile category who originated in the current quarter. \% Inquired indicates the fraction of borrowers within each credit risk profile category who made an inquiry in the current quarter. \% Originated (Borrowers with Inquiries) reflects the fraction of borrowers who had an inquiry in the current or previous quarter and subsequently originated in the current quarter. \% of Borrowers with reports the percentage of borrowers in each credit risk category who have a first mortgage or credit limit. For example, 1.36\% of borrowers with a non-zero credit limit are classified as deep subprime borrowers, even though they represent 6\% of the population. Therefore, only 22.7\% of deep subprime borrowers have a non-zero credit limit. The table reports credit limits, first mortgage balances, and their ratios to household income for those with non-zero credit card limits and first mortgage balances. Source: Authors' calculations based on Experian Data.
\end{tablenotes}
\end{threeparttable}
\end{table}

%% file: Tables/credit_risk_reg.tex
\begin{sidewaystable}[htbp]\centering
\begin{threeparttable}
\scriptsize
\def\sym#1{\ifmmode^{#1}\else\(^{#1}\)\fi}
\caption{Access to Credit by Risk Profile \label{tab:crr_reg}}
\begin{tabular*}{\hsize}{@{\hskip\tabcolsep\extracolsep\fill}l*{8}{p{0.7in}}}
\hline\hline
& \multicolumn{4}{c}{Credit Cards} & \multicolumn{4}{c}{Mortgages} \\
& & & & & & & & \\[\dimexpr-\normalbaselineskip+2pt]
                    &\multicolumn{1}{c}{(1)}&\multicolumn{1}{c}{(2)}&\multicolumn{1}{c}{(3)}&\multicolumn{1}{c}{(4)}&\multicolumn{1}{c}{(5)}&\multicolumn{1}{c}{(6)}&\multicolumn{1}{c}{(7)}&\multicolumn{1}{c}{(8)}\\
                    &\multicolumn{1}{c}{Credit Limit} &\multicolumn{1}{c}{Inquiries}&\multicolumn{1}{c}{ Origination}&\multicolumn{1}{c}{Originations if} &\multicolumn{1}{c}{Credit Limit} &\multicolumn{1}{c}{Inquiries}&\multicolumn{1}{c}{ Origination}&\multicolumn{1}{c}{Originations if} \\
                    &\multicolumn{1}{c}{/ Balances} &\multicolumn{1}{c}{} &\multicolumn{1}{c}{} &\multicolumn{1}{c}{Inquired} &\multicolumn{1}{c}{/ Balances} &\multicolumn{1}{c}{} &\multicolumn{1}{c}{} &\multicolumn{1}{c}{Inquired} \\
\hline
& & & & & & & & \\[\dimexpr-\normalbaselineskip+2pt]
Prime        &  -7917.347&      0.037&      0.011&     -0.011 &  20126.692&      0.033&      0.007&     -0.031\\
                    &  (130.652)         &    (0.001)         &    (0.001)         &    (0.005)         &  (494.724)         &    (0.001)         &    (0.001)         &    (0.005)         \\
& & & & & & & & \\[\dimexpr-\normalbaselineskip+2pt]
Near Prime        & -14383.424&      0.115&      0.015&     -0.047&   9405.433&      0.073&      0.010&     -0.055\\
                    &  (173.008)         &    (0.001)         &    (0.002)         &    (0.006)         &  (646.995)         &    (0.001)         &    (0.001)         &    (0.005)         \\
& & & & & & & & \\[\dimexpr-\normalbaselineskip+2pt]
Subprime        & -28332.291&      0.143&     -0.071&     -0.189& -18646.570&      0.050&     -0.003&     -0.093\\
                    &  (121.642)         &    (0.001)         &    (0.001)         &    (0.005)         &  (420.904)         &    (0.001)         &    (0.001)         &    (0.005)         \\
& & & & & & & & \\[\dimexpr-\normalbaselineskip+2pt]
Deep Subprime        & -31015.469&      0.183&     -0.109&     -0.282& -24759.575&      0.040&     -0.010&     -0.120\\
                    &  (127.105)         &    (0.002)         &    (0.001)         &    (0.005)         &  (505.414)         &    (0.001)         &    (0.001)         &    (0.005)         \\
& & & & & & & & \\[\dimexpr-\normalbaselineskip+2pt]
30-39               &   9685.490&      0.008&      0.011&     -0.001         &  48881.809&      0.037&      0.009&     -0.002\\
                    &   (70.860)         &    (0.000)         &    (0.000)         &    (0.001)         &  (417.223)         &    (0.000)         &    (0.000)         &    (0.001)         \\
& & & & & & & & \\[\dimexpr-\normalbaselineskip+2pt]
40-49               &  15675.728&      0.002&      0.013&     -0.005&  65633.694&      0.032&      0.008&     -0.009\\
                    &  (108.016)         &    (0.000)         &    (0.000)         &    (0.001)         &  (608.152)         &    (0.000)         &    (0.000)         &    (0.001)         \\
& & & & & & & & \\[\dimexpr-\normalbaselineskip+2pt]
50-59               &  18744.131&     -0.009&      0.009&     -0.012&  53565.830&      0.019&      0.005&     -0.013\\
                    &  (123.438)         &    (0.000)         &    (0.000)         &    (0.001)         &  (601.340)         &    (0.000)         &    (0.000)         &    (0.001)         \\
& & & & & & & & \\[\dimexpr-\normalbaselineskip+2pt]
60-69               &  17673.841&     -0.021&     -0.001 &     -0.013&  32022.655&      0.006&      0.001&     -0.018\\
                    &  (134.698)         &    (0.000)         &    (0.000)         &    (0.001)         &  (541.962)         &    (0.000)         &    (0.000)         &    (0.001)         \\
& & & & & & & & \\[\dimexpr-\normalbaselineskip+2pt]
70+                 &   7636.702&     -0.042&     -0.031&     -0.021&    -30.810         &     -0.015&     -0.006&     -0.033\\
                    &  (123.213)         &    (0.000)         &    (0.000)         &    (0.001)         &  (404.867)         &    (0.000)         &    (0.000)         &    (0.001)         \\
                    
& & & & & & & & \\[\dimexpr-\normalbaselineskip+2pt]
Constant            & -11502.599&      0.062&      0.087&      0.362&  11335.228&      0.014&      0.013&      0.159\\
                    &  (314.503)         &    (0.001)         &    (0.001)         &    (0.003)         &  (526.828)         &    (0.001)         &    (0.000)         &    (0.004)         \\
\hline
& & & & & & & & \\[\dimexpr-\normalbaselineskip+2pt]
Observations        &26,147,712      &26,147,712      &26,147,712     &3,761,219      &26,147,712      &26,147,712      &26,147,712     &2,235,063      \\
$R^2$               &      0.2518         &      0.0305         &      0.0367         &      0.0705       &      0.1241         &      0.0208         &      0.0073         &      0.0298         \\ 
\hline\hline
\end{tabular*}
\begin{tablenotes}
\footnotesize
\item Notes: This table reports estimated for equation \ref{reg:access_spec} for credit cards (columns 1 to 4), and mortgages (columns 5 to 8). Column (1) is credit limits for credit cards, while Column (5) is first mortgage balances for mortgages. Column (4) and (8) restrict the sample to individuals who inquired in the current or previous quarter and estimates the coefficients for origination. Credit risk profiles based on previous quarter. Standard errors are clustered by ZIP code.
\end{tablenotes}
\end{threeparttable}
\end{sidewaystable}

%% file: Tables/vulnerable_desc.tex
\begin{sidewaystable}[htbp]\centering
\begin{threeparttable}
\footnotesize
\def\sym#1{\ifmmode^{#1}\else\(^{#1}\)\fi}
\caption{Descriptive Statistics (Vulnerable Populations)}\label{tab:desc_vulnerable}
\begin{tabular}[c]{lccccccccc}
\hline\hline
& & & & & & & & &  \\[\dimexpr-\normalbaselineskip+2pt]
            &\multicolumn{1}{c}{Age: $<$ 30}&\multicolumn{1}{c}{Age $\ge$ 30}&\multicolumn{1}{c}{Diff$_{Age}$}&\multicolumn{1}{c}{Income$_{p20}$}&\multicolumn{1}{c}{Income$_{>p20}$}&\multicolumn{1}{c}{Diff$_{Income}$}&\multicolumn{1}{c}{Minority}&\multicolumn{1}{c}{Non-Minority}&\multicolumn{1}{c}{Diff$_{Minority}$}\\ \hline
& & & & & & & & &  \\[\dimexpr-\normalbaselineskip+2pt]
\multicolumn{10}{l}{\underline{Panel A: Experian Sample}} \\ 
& & & & & & & & &  \\ 
Default     &        0.23&        0.17&       -0.06\sym{***}&        0.37&        0.14&       -0.23\sym{***}&        0.28&        0.16&       -0.12\sym{***}\\
            &      (0.42)&      (0.38)&                     &      (0.48)&      (0.35)&                     &      (0.45)&      (0.37)&                     \\
Model Percentile&       40.46&       52.65&       12.19\sym{***}&       26.33&       55.60&       29.27\sym{***}&       39.23&       52.46&       13.22\sym{***}\\
            &     (24.13)&     (29.87)&                     &     (18.62)&     (28.38)&                     &     (27.52)&     (28.96)&                     \\
Credit Score Percentile&       38.68&       53.14&       14.47\sym{***}&       22.70&       56.46&       33.76\sym{***}&       38.54&       52.61&       14.08\sym{***}\\
            &     (24.15)&     (29.65)&                     &     (18.20)&     (27.48)&                     &     (27.50)&     (28.89)&                     \\
Total Balances     &       22.63&       85.08&       62.45\sym{***}&       10.15&       86.02&       75.87\sym{***}&       48.66&       76.71&       28.05\sym{***}\\
            &     (65.33)&    (183.69)&                     &     (39.37)&    (182.08)&                     &    (122.40)&    (175.56)&                     \\
90 DPD+ Balances     &        1.25&        3.48&        2.23\sym{***}&        1.99&        3.23&        1.24\sym{***}&        4.32&        2.69&       -1.63\sym{***}\\
            &     (14.03)&     (36.10)&                     &     (15.06)&     (35.51)&                     &     (35.51)&     (31.91)&                     \\
Household Income&       46.95&       87.05&       40.10\sym{***}&       29.68&       89.84&       60.16\sym{***}&       62.54&       81.93&       19.40\sym{***}\\
            &     (32.93)&     (56.81)&                     &      (6.54)&     (55.23)&                     &     (43.39)&     (56.82)&                     \\
Age         &       23.81&       51.56&       27.76\sym{***}&       32.70&       48.57&       15.87\sym{***}&       43.72&       45.94&        2.22\sym{***}\\
            &      (3.31)&     (13.71)&                     &     (13.17)&     (16.06)&                     &     (16.15)&     (16.86)&                     \\
\hline
& & & & & & & & &  \\[\dimexpr-\normalbaselineskip+2pt]
\(N\)       &     5,704,564&    20,536,315&    26,240,879    &     5,021,250&    21,219,629&    26,240,879&         4,873,733&    21,367,146&    26,240,879                    \\ \hline
& & & & & & & & &  \\ 
\multicolumn{10}{l}{\underline{Panel B: HMDA Matched Sample}} \\ 
& & & & & & & & &  \\ 
Default     &        0.13&        0.11&       -0.01\sym{***}&        0.22&        0.10  &     -0.12\sym{***}&        0.19&        0.10    &   -0.09\sym{***}\\
            &      (0.33)&      (0.31)&                     &      (0.42)&      (0.30)  &                   &      (0.39)&      (0.30)    &                 \\
Model Percentile&       44.22&       52.94&        8.73\sym{***}&       30.75&       53.62  &     22.88\sym{***}&       40.22&       52.98   &    12.76\sym{***}\\
            &     (21.45)&     (26.17)&                     &     (18.30)&     (24.94)  &                   &     (24.71)&     (24.94)   &                  \\
Credit Score Percentile&       47.37&       59.06&       11.68\sym{***}&       33.13&       59.43 &      26.30\sym{***}&       46.68&       58.13    &   11.46\sym{***}\\
            &     (21.45)&     (25.08)&                     &     (18.72)&     (23.74) &                    &     (24.62)&     (24.31)   &                  \\
Total Balances     &       44.52&      158.61&      114.09\sym{***}&       23.59&      145.42   &   121.83\sym{***}&      104.03&      135.73     &  31.70\sym{***}\\
            &     (83.85)&    (187.30)&                     &     (59.03)&    (180.39) &                    &    (139.25)&    (180.83)   &                  \\
90 DPD+ Balances    &        0.97&        2.39&        1.42\sym{***}&        1.57&        2.11    &    0.54\sym{***}&        3.38&        1.76    &   -1.62\sym{***}\\
            &     (14.25)&     (26.35)&                     &     (15.30)&     (24.91) &                    &     (29.96)&     (22.45)   &                  \\
Household Income&       50.82&       99.56&       48.75\sym{***}&       30.30&       95.57   &    65.27\sym{***}&       71.37&       90.81    &   19.44\sym{***}\\
            &     (30.85)&     (57.67)&                     &      (6.45)&     (55.64)  &                   &     (44.36)&     (58.03)   &                  \\
Age         &       24.32&       46.87&       22.55\sym{***}&       28.04&       43.15    &   15.10\sym{***}&       40.07&       41.52    &    1.45\sym{***}\\
            &      (3.21)&     (11.51)&                     &      (9.82)&     (13.53) &                    &     (13.39)&     (14.16)   &                  \\
\hline
& & & & & & & & &  \\[\dimexpr-\normalbaselineskip+2pt]
\(N\)       &      188,005&      568,174&      756,179         &       94,176&      662,003&      756,179                     &      130,869&      625,310&      756,179                    \\

\hline\hline
\end{tabular}
\begin{tablenotes}
\footnotesize
\item  Notes: \sym{***} denotes statistical significance at the 1\% level. Differences in means between subgroups are tested via a t-test (results in ``Diff" columns). Income and debt figures are in thousands. In Panel A, 'Minority status' is based on a ZIP code's Black and Hispanic population being over 50\%. In Panel B, it is assigned if an individual is Black or Hispanic.
\end{tablenotes}
\end{threeparttable}
\end{sidewaystable}

%% file: Tables/vulnerable_nonprime.tex
\begin{sidewaystable}[htbp]
\begin{center}
\begin{threeparttable}
\scriptsize
\def\sym#1{\ifmmode^{#1}\else\(^{#1}\)\fi}
\caption{Vulnerable Populations: Non-Prime Borrowers \label{tab:vulnerable_populations_nonprime}}
\begin{tabular*}{\hsize}{@{\hskip\tabcolsep\extracolsep\fill}p{1.1in}*{8}{p{0.55in}}}
\hline\hline
& & & & & & & &\\[\dimexpr-\normalbaselineskip+2pt]
                    &\multicolumn{1}{c}{(1)}&\multicolumn{1}{c}{(2)}&\multicolumn{1}{c}{(3)}&\multicolumn{1}{c}{(4)}&\multicolumn{1}{c}{(5)}&\multicolumn{1}{c}{(6)}&\multicolumn{1}{c}{(7)}&\multicolumn{1}{c}{(8)}\\
\multicolumn{9}{l}{DV: Our Model Ranking - Credit Score Ranking}\\
\hline
& & & & & & & &\\[\dimexpr-\normalbaselineskip+2pt]
 \multicolumn{9}{l}{\underline{Panel A: Experian Sample}} \\ 
& & & & & & & &\\[\dimexpr-\normalbaselineskip+2pt]
Default &                     &    -10.616\sym{***}&                     &    -11.317\sym{***}&                     &    -10.699\sym{***}&                     &    -10.612\sym{***}\\
                    &                     &    (0.040)         &                     &    (0.045)         &                     &    (0.047)         &                     &    (0.037)         \\
& & & & & & & & \\[\dimexpr-\normalbaselineskip+2pt]
Income$_{p20}$  &                     &                     &      3.367\sym{***}&      3.444\sym{***}&                     &                     &                     &                     \\
                    &                     &                     &    (0.062)         &    (0.056)         &                     &                     &                     &                     \\
& & & & & & & & \\[\dimexpr-\normalbaselineskip+2pt]
Income$_{p20}$ $\times$ Default &                     &                     &                     &      1.234\sym{***}&                     &                     &                     &                     \\
                    &                     &                     &                     &    (0.033)         &                     &                     &                     &                     \\
& & & & & & & & \\[\dimexpr-\normalbaselineskip+2pt]
Young            &                     &                     &                     &                     &      2.974\sym{***}&      2.591\sym{***}&                     &                     \\
                    &                     &                     &                     &                     &    (0.080)         &    (0.086)         &                     &                     \\
& & & & & & & & \\[\dimexpr-\normalbaselineskip+2pt]
Young $\times$ Default &                     &                     &                     &                     &                     &      0.452\sym{***}&                     &                     \\
                    &                     &                     &                     &                     &                     &    (0.061)         &                     &                     \\
& & & & & & & & \\[\dimexpr-\normalbaselineskip+2pt]
Minority           &                     &                     &                     &                     &                     &                     &      0.216\sym{***}&      0.942\sym{***}\\
                    &                     &                     &                     &                     &                     &                     &    (0.029)         &    (0.032)         \\
& & & & & & & & \\[\dimexpr-\normalbaselineskip+2pt]
Minority $\times$ Default &                     &                     &                     &                     &                     &                     &                     &     -0.157\sym{***}\\
                    &                     &                     &                     &                     &                     &                     &                     &    (0.035)         \\
& & & & & & & & \\[\dimexpr-\normalbaselineskip+2pt]
Constant            &      3.555\sym{***}         &      7.787\sym{***}&      2.252\sym{***}&      6.528\sym{***}&      2.707\sym{***}&      7.030\sym{***}&      3.498\sym{***}&      7.555\sym{***}\\
                    &         (0.000)         &    (0.016)         &    (0.024)         &    (0.023)         &    (0.023)         &    (0.029)         &    (0.008)         &    (0.015)         \\
 \hline  
& & & & & & & &\\[\dimexpr-\normalbaselineskip+2pt]
Observations        &10,508,324       &10,508,324       &10,508,324       &10,508,324       &10,508,324       &10,508,324       &10,508,324       &10,508,324       \\
$R^2$               &      0.0042         &      0.1199         &      0.0157         &      0.1360         &      0.0119         &      0.1266         &      0.0042         &      0.1204         \\
\hline 
& & & & & & & &\\[\dimexpr-\normalbaselineskip+2pt]
\multicolumn{9}{l}{\underline{Panel B: Experian-HMDA Sample}} \\
& & & & & & & & \\[\dimexpr-\normalbaselineskip+2pt]
Default     &                     &     -7.197\sym{***}&                     &     -7.091\sym{***}&                     &     -5.958\sym{***}&                     &     -7.629\sym{***}\\
                    &                     &    (0.066)         &                     &    (0.071)         &                     &    (0.064)         &                     &    (0.074)         \\
& & & & & & & & \\[\dimexpr-\normalbaselineskip+2pt]
Income$_{p20}$    &                     &                     &      2.950\sym{***}&      3.028\sym{***}&                     &                     &                     &                     \\
                    &                     &                     &    (0.105)         &    (0.126)         &                     &                     &                     &                     \\
& & & & & & & & \\[\dimexpr-\normalbaselineskip+2pt]
Income$_{p20}$ $\times$ Default &                     &                     &                     &     -0.341\sym{**} &                     &                     &                     &                     \\
                    &                     &                     &                     &    (0.134)         &                     &                     &                     &                     \\
& & & & & & & & \\[\dimexpr-\normalbaselineskip+2pt]
Young                  &                     &                     &                     &                     &      4.348\sym{***}&      4.846\sym{***}&                     &                     \\
                    &                     &                     &                     &                     &    (0.094)         &    (0.112)         &                     &                     \\
& & & & & & & & \\[\dimexpr-\normalbaselineskip+2pt]
Young $\times$ Default &                     &                     &                     &                     &                     &     -3.262\sym{***}&                     &                     \\
                    &                     &                     &                     &                     &                     &    (0.128)         &                     &                     \\
& & & & & & & & \\[\dimexpr-\normalbaselineskip+2pt]
Minority           &                     &                     &                     &                     &                     &                     &     -1.811\sym{***}&     -1.926\sym{***}\\
                    &                     &                     &                     &                     &                     &                     &    (0.064)         &    (0.085)         \\
& & & & & & & & \\[\dimexpr-\normalbaselineskip+2pt]
Minority $\times$ Default &                     &                     &                     &                     &                     &                     &                     &      1.791\sym{***}\\
                    &                     &                     &                     &                     &                     &                     &                     &    (0.118)         \\
& & & & & & & & \\[\dimexpr-\normalbaselineskip+2pt]
Constant            &     -1.766\sym{***}&      0.496\sym{***}&     -2.588\sym{***}&     -0.352\sym{***}&     -3.173\sym{***}&     -1.170\sym{***}&     -1.291\sym{***}&      0.963\sym{***}\\
                    &    (0.000)         &    (0.021)         &    (0.029)         &    (0.038)         &    (0.031)         &    (0.039)         &    (0.017)         &    (0.029)         \\
\hline
& & & & & & & &\\[\dimexpr-\normalbaselineskip+2pt]
Observations        & 217,516       & 217,516       & 217,516       & 217,516       & 217,516       & 217,516       & 217,516       & 217,516       \\
$R^2$               &      0.0320         &      0.0938         &      0.0416         &      0.1032         &      0.0542         &      0.1140         &      0.0354         &      0.0962         \\
\hline\hline
\end{tabular*}
\begin{tablenotes}
\scriptsize
\item Notes: This table reports regressions of the form specified by Equation (\ref{reg:spec}). The dependent variable is the percentile ranking difference between our model and the credit score. The independent variable, "Default", is a binary indicator showing whether the individual defaulted within the next two years. The variable "Young" is a binary indicator for individuals under 30 years old. The variable "Income$_{p20}$" indicates whether the individual is in the lowest quintile of the estimated income distribution. In Panel A, the Minority indicator is based on ZIP code level Black and Hispanic population being over 50\%. In Panel B, the Minority indicator is 1 if the individual is Black or Hispanic. Panel A presents the full Experian sample, while Panel B provides the matched Experian-HMDA sample results. We control for State x Quarter fixed effects. Standard errors are clustered by state and quarter. \sym{*} \(p<0.10\), \sym{**} \(p<0.05\), \sym{***} \(p<0.01\).
\end{tablenotes}
\end{threeparttable}
\end{center}
\end{sidewaystable}

%% file: Tables/vulnerable_prime.tex
\begin{sidewaystable}[htbp]
\begin{center}
\begin{threeparttable}
\scriptsize
\def\sym#1{\ifmmode^{#1}\else\(^{#1}\)\fi}
\caption{Vulnerable Populations: Prime Borrowers \label{tab:vulnerable_populations_prime}}
\begin{tabular*}{\hsize}{@{\hskip\tabcolsep\extracolsep\fill}p{1.1in}*{8}{p{0.55in}}}
\hline\hline
& & & & & & & &\\[\dimexpr-\normalbaselineskip+2pt]
                    &\multicolumn{1}{c}{(1)}&\multicolumn{1}{c}{(2)}&\multicolumn{1}{c}{(3)}&\multicolumn{1}{c}{(4)}&\multicolumn{1}{c}{(5)}&\multicolumn{1}{c}{(6)}&\multicolumn{1}{c}{(7)}&\multicolumn{1}{c}{(8)}\\
\multicolumn{9}{l}{DV: Our Model Ranking - Credit Score Ranking}\\
\hline
& & & & & & & &\\[\dimexpr-\normalbaselineskip+2pt]
 \multicolumn{9}{l}{\underline{Panel A: Experian Sample}} \\ 
& & & & & & & &\\[\dimexpr-\normalbaselineskip+2pt]
Default    &                     &    -11.031\sym{***}&                     &    -11.249\sym{***}&                     &    -11.823\sym{***}&                     &    -10.914\sym{***}\\
                    &                     &    (0.099)         &                     &    (0.110)         &                     &    (0.120)         &                     &    (0.090)         \\
& & & & & & & & \\[\dimexpr-\normalbaselineskip+2pt]
Income$_{p20}$   &                     &                     &     -2.550\sym{***}&     -2.256\sym{***}&                     &                     &                     &                     \\
                    &                     &                     &    (0.112)         &    (0.115)         &                     &                     &                     &                     \\
& & & & & & & & \\[\dimexpr-\normalbaselineskip+2pt]
Income$_{p20}$ $\times$ Default &                     &                     &                     &      2.950\sym{***}&                     &                     &                     &                     \\
                    &                     &                     &                     &    (0.131)         &                     &                     &                     &                     \\
& & & & & & & & \\[\dimexpr-\normalbaselineskip+2pt]
Young           &                     &                     &                     &                     &     -0.135\sym{**} &     -0.138\sym{**} &                     &                     \\
                    &                     &                     &                     &                     &    (0.064)         &    (0.062)         &                     &                     \\
& & & & & & & & \\[\dimexpr-\normalbaselineskip+2pt]
Young $\times$ Default &                     &                     &                     &                     &                     &      3.513\sym{***}&                     &                     \\
                    &                     &                     &                     &                     &                     &    (0.105)         &                     &                     \\
& & & & & & & & \\[\dimexpr-\normalbaselineskip+2pt]
Minority            &                     &                     &                     &                     &                     &                     &     -0.731\sym{***}&     -0.370\sym{***}\\
                    &                     &                     &                     &                     &                     &                     &    (0.036)         &    (0.037)         \\
& & & & & & & & \\[\dimexpr-\normalbaselineskip+2pt]
Minority $\times$ Default &                     &                     &                     &                     &                     &                     &                     &     -0.386\sym{***}\\
                    &                     &                     &                     &                     &                     &                     &                     &    (0.091)         \\
& & & & & & & & \\[\dimexpr-\normalbaselineskip+2pt]
Constant            &     -2.373\sym{***}&     -1.924\sym{***}&     -2.218\sym{***}&     -1.794\sym{***}&     -2.350\sym{***}&     -1.901\sym{***}&     -2.276\sym{***}&     -1.876\sym{***}\\
                    &    (0.000)         &    (0.004)         &    (0.007)         &    (0.008)         &    (0.011)         &    (0.011)         &    (0.005)         &    (0.005)         \\
 \hline  
& & & & & & & &\\[\dimexpr-\normalbaselineskip+2pt]
Observations        &15,732,555        &15,732,555        &15,732,555        &15,732,555        &15,732,555        &15,732,555        &15,732,555        &15,732,555        \\
$R^2$               &      0.0029         &      0.0227         &      0.0045         &      0.0238         &      0.0029         &      0.0230         &      0.0031         &      0.0227         \\
\hline 
& & & & & & & &\\[\dimexpr-\normalbaselineskip+2pt]
\multicolumn{9}{l}{\underline{Panel B: Experian-HMDA Sample}} \\
& & & & & & & & \\[\dimexpr-\normalbaselineskip+2pt]
Default       &                     &     -7.080\sym{***}&                     &     -7.146\sym{***}&                     &     -7.369\sym{***}&                     &     -6.886\sym{***}\\
                    &                     &    (0.111)         &                     &    (0.119)         &                     &    (0.128)         &                     &    (0.126)         \\
& & & & & & & & \\[\dimexpr-\normalbaselineskip+2pt]
Income$_{p20}$  &                     &                     &     -1.335\sym{***}&     -1.227\sym{***}&                     &                     &                     &                     \\
                    &                     &                     &    (0.103)         &    (0.106)         &                     &                     &                     &                     \\
& & & & & & & & \\[\dimexpr-\normalbaselineskip+2pt]
Income$_{p20}$ $\times$ Default &                     &                     &                     &      1.128\sym{***}&                     &                     &                     &                     \\
                    &                     &                     &                     &    (0.340)         &                     &                     &                     &                     \\
& & & & & & & & \\[\dimexpr-\normalbaselineskip+2pt]
Young                 &                     &                     &                     &                     &      0.773\sym{***}&      0.726\sym{***}&                     &                     \\
                    &                     &                     &                     &                     &    (0.094)         &    (0.094)         &                     &                     \\
& & & & & & & & \\[\dimexpr-\normalbaselineskip+2pt]
Young $\times$ Default &                     &                     &                     &                     &                     &      1.298\sym{***}&                     &                     \\
                    &                     &                     &                     &                     &                     &    (0.250)         &                     &                     \\
& & & & & & & & \\[\dimexpr-\normalbaselineskip+2pt]
Minority      &                     &                     &                     &                     &                     &                     &     -2.085\sym{***}&     -1.918\sym{***}\\
                    &                     &                     &                     &                     &                     &                     &    (0.050)         &    (0.050)         \\
& & & & & & & & \\[\dimexpr-\normalbaselineskip+2pt]
Minority $\times$ Default &                     &                     &                     &                     &                     &                     &                     &     -0.219         \\
                    &                     &                     &                     &                     &                     &                     &                     &    (0.235)         \\
& & & & & & & & \\[\dimexpr-\normalbaselineskip+2pt]
Constant            &     -6.833\sym{***}         &     -6.589\sym{***}&     -6.750\sym{***}&     -6.514\sym{***}&     -7.001\sym{***}&     -6.747\sym{***}&     -6.547\sym{***}&     -6.331\sym{***}\\
                    &         (0.000)         &    (0.004)         &    (0.006)         &    (0.007)         &    (0.021)         &    (0.020)         &    (0.007)         &    (0.008)         \\
 \hline
& & & & & & & &\\[\dimexpr-\normalbaselineskip+2pt]
Observations        & 538,663       & 538,663       & 538,663       & 538,663       & 538,663       & 538,663       & 538,663       & 538,663       \\
$R^2$               &      0.0523         &      0.0636         &      0.0530         &      0.0641         &      0.0530         &      0.0643         &      0.0556         &      0.0664         \\
\hline\hline
\end{tabular*}
\begin{tablenotes}
\scriptsize
\item Notes: This table reports regressions of the form specified by Equation (\ref{reg:spec}). The dependent variable is the percentile ranking difference between our model and the credit score. The independent variable, "Default", is a binary indicator showing whether the individual defaulted within the next two years. The variable "Young" is a binary indicator for individuals under 30 years old. The variable "Income$_{p20}$" indicates whether the individual is in the lowest quintile of the estimated income distribution. In Panel A, the Minority indicator is based on ZIP code level Black and Hispanic population being over 50\%. In Panel B, the Minority indicator is 1 if the individual is Black or Hispanic. Panel A presents the full Experian sample, while Panel B provides the matched Experian-HMDA sample results. We control for State x Quarter fixed effects. Standard errors are clustered by state and quarter. \sym{*} \(p<0.10\), \sym{**} \(p<0.05\), \sym{***} \(p<0.01\).
\end{tablenotes}
\end{threeparttable}
\end{center}
\end{sidewaystable}

%% file: Sections/hmda_match.tex
\section*{Online Appendix: Matched Experian-HMDA Panel}\label{app:hmda_match}

\subsection*{HMDA Data}
The Home Mortgage Disclosure Act mandates that almost all mortgage lenders report comprehensive information on the applications they receive and whether they approve the loan.\footnote{Exceptions are granted only for very small or strictly rural lenders (\citeN{butler2022racial}). Approximately 95\% of all first-lien mortgages are reported to the HMDA database (\citeN{avery2017profile}), with the coverage rate likely higher for properties in MSAs.} Mortgage lenders submit applicants' racial and ethnic background, personal attributes, and loan application details, such as the requested loan size, income, loan purpose (purchase, refinancing, improvement), co-applicants, loan priority (first or second lien), and the census tract location of the property, to the HMDA database. If a loan is granted, any loan sale is reported, along with an indicator for sale to any quasi-government entity. Our HMDA data ranges from 2010 to 2017.

\input{Tables/sample_restrictions_hmda}

\subsection*{Credit Bureau - HMDA Matching}

Both the credit bureau and HMDA data are anonymized and lack a unique identifier to directly link the two datasets. However, the detailed information on originated mortgages in both datasets facilitates the matching of mortgages based on their characteristics, as outlined by \citeN{butler2022racial}. Following \citeN{butler2022racial}, we implement a series of filters on the mortgages extracted from our HMDA and credit bureau data. For HMDA, we ensure each mortgage is a first lien and located within a Metropolitan Statistical Area (MSA), where HMDA data is most comprehensive. The borrower must reside in one of the 50 states, including Washington D.C., following loan origination. We focus on home purchase and refinancing loans, excluding home improvement loans due to their less defined nature in both datasets. Additionally, we limit our analysis to owner-occupied properties to align the property location with the borrower's location in the credit bureau data and consider only mortgages with a single applicant or borrower, ensuring that the demographic data is directly relevant to the individual matched in the credit bureau data. We apply similar filters to our credit bureau data. We consider only first-lien, non-joint mortgages, where balances for both types of mortgages are available in the Experian data. The residential ZIP code is restricted to those covered in our final HMDA sample, ensuring it falls within an MSA in one of the 50 states or Washington D.C. Moreover, to enhance the accuracy of matching balances in refinancing cases and increase the likelihood of identifying owner-occupied properties, we restrict our sample to individuals with only a single first mortgage trade in the current quarter and at most one in the previous two quarters. This careful selection process aims to enhance the accuracy of our analysis. For itemized sample restrictions, see Table \ref{tab:sample_restrictions_experian}. 

After applying the filters to both the HMDA and credit bureau data, the target population for the matched sample comprises borrowers obtaining a first-lien home purchase or refinance loan individually (no co-applicant), for their primary residence located within an MSA in the United States excluding its territories, between 2010 and 2017.

Following the procedure outlined in \citeN{butler2022racial} as a reference, we made a slight modification in our matching process by substituting the census tract with ZCTA5 as in \citeN{lavoice2024racial}. Specifically, we matched mortgages in the credit bureau data to HMDA data based on six characteristics: origination year, ZCTA5, loan amount, loan purpose (purchase or refinancing), mortgage type (conventional, FHA, or VA), and if/to which quasi-government entity the loan is sold (purchaser type).\footnote{These six variables are directly observable in the HMDA data. We provide additional details and validation for our credit bureau data in the Appendix \ref{appendixa}} We implement an exact matching algorithm, as opposed to using nearest matches or propensity scores, to ensure the highest accuracy for our matched dataset. 89.73\% of originated mortgages in the HMDA data are unique when using our six matching variables, with ZCTA5 being replaced by the census tract. We refine our HMDA dataset by removing duplicates based on these six matching variables, where we use the census tract instead of ZCTA5.\footnote{To address the issue of integer-based origination balances in the HMDA data, we adopt a range-based approach for origination balances in the credit bureau data. This method, which includes adjacent integer values, helps to avoid inaccuracies stemming from rounding differences. For all other matching variables, we ensure precise, exact matches.} 

Because our Experian credit report data only reports an individual's zip code, we first matched our zip code level credit bureau data to Census 5-Digit Zip Code Tabulation Areas (ZCTA5s) using the Missouri Census Data Center's Census Tract - ZIP/ZCTA crosswalk files.\footnote{USPS zip codes are not areal features used by the Census but a collection of mail delivery routes that identify the individual post office or metropolitan area delivery station associated with mailing addresses. ZIP Code Tabulation Areas (ZCTAs) are generalized areal representations of United States Postal Service (USPS) ZIP Code service areas. HMDA data reports 2010 census tracts starting in 2013. For loans originated from 2003 to 2012, HMDA data uses the 2000 census tracts. We successfully matched ZCTAs for 95.5\% of Census Tracts observations in the HMDA data. Crosswalk URL:  \url{https://mcdc.missouri.edu/applications/geocorr2014.html}.} This process increased our HMDA sample from 19.33M to 35.44M observations as it assigns each individual to all potential ZCTA5s. We observe a one-to-one match in 51.6\% of cases, a one-to-two match in 29.2\%, a one-to-three match in 11.01\%, a one-to-four match in 4.31\%, and 3.88\% of cases involve more than four ZCTAs. However, duplication issues arise, particularly when two (or more) loans differ only at the census tract level but are assigned to the same ZCTA5. Upon expanding our dataset to the ZCTA5 level, we engage in a second phase of duplicate removal, this time targeting the matching variable level. This results in a dataset comprising 27.4 million observations. Notably, this process impacts 32.76\% of the original census tract level observations. By implementing this two-stage duplicate elimination approach — initially at the census tract level and subsequently at the ZCTA5 level post-expansion — we reduce the incidence of false positives, enhancing the reliability of our data.

\citeN{butler2022racial} identifies two potential sources of error when matching HMDA with credit bureau data. First, a data error in one of the matching variables might lead to a mismatch. Second, HMDA-reporting and non-reporting lenders may originate identical, otherwise unique mortgages. They argue that such mismatches should not introduce bias in estimates beyond pure noise. 

In our matching process, expanding HMDA data from the census tract to ZCTA5 level introduces two potential errors. The first concerns assigning individuals from a specific census tract to multiple ZIP codes, raising the chance of matching with an incorrect ZCTA5. The risk is partially offset by two factors: first, the high degree of uniqueness in the HMDA data at the census tract level, evidenced by 89.73\% of mortgages being distinct. Second, the comprehensive coverage of the HMDA dataset, which includes most originated mortgages, significantly enhances the likelihood of accurately identifying the correct loan when mapping to the ZCTA5 level. The second issue arises when loans from different census tracts, sharing identical matching variables, are grouped under the same ZCTA5.\footnote{To view this from another angle, we assigned unique pseudo-IDs to each loan in the census tract-level HMDA data and then expanded these to the ZCTA5 level. We found that 67.24\% of these pseudo-IDs had no conflicts at the ZCTA5 level. However, conflicts did arise: 23.57\% of pseudo-IDs conflicted with one other, 6.62\% with two others, and 2.58\% with more than two. These conflicts affected 32.76\% of HMDA observations at the census tract level, leading to the removal of 18.47\% of pseudo-IDs. Our approach does not involve removing all conflicted pseudo-IDs; rather, we only eliminate those in ZIP codes with identified conflicts. This means a single census tract could span two ZCTAs, with conflicts present in one but not the other. We retain conflict-free pseudo-IDs. However, excluding ZIP codes with conflicts reduces the efficiency of our matching process if the correct ZIP code is among those removed.} By eliminating duplicates at the match variable level, we address this concern.

We report the match rate as well as summary statistics on the match in Table \ref{table:1}. Panel A displays the match rate for home purchase mortgages, refinance loans, and both types of loans combined. We find a corresponding HMDA mortgage for 43.02\% of the mortgages in the credit bureau data with a slightly higher match rate for refinance loans as opposed to home purchase loans. Given that HMDA covers 95\% of mortgages, 90\% are unique at the census tract level, 95.5\% of census tracts are merged to ZCTA5, and 67.24\% of census tract level are unaffected by the ZCTA5 expansion, the conservative lower bound on the best achievable match rate is the product of these four numbers, approximately 54.9\%.\footnote{Our estimated lower bound for the success rate of matches at the ZCTA5 level stands at 64.19\% (0.955*0.6724) relative to the potential match rate achievable at the more precise census-tract level. Conversely, the upper bound for this success rate is 77.86\% (0.955*0.8153). This lower bound is based on the assumption that in all cases of conflict, it was always the conflicting loans that would have been the correct matches. On the other hand, the upper bound is calculated under the assumption that the correct loans were never among those deleted, thereby allowing us to achieve a higher rate of successful matches.} In other words, our algorithm successfully found matches in roughly 78.4\% (0.4302/0.549) of the potential cases. 

The summary statistics in panels B and C of Table \ref{table:1} assess whether our matched sample is representative of the original population for home purchase loans and refinance loans respectively. Panel B reveals that the successfully matched home purchase mortgage sample is largely representative of the initial population of credit bureau mortgages. Similarly, Panel C demonstrates that the matched sample of refinance loans accurately represents the starting sample from the credit bureau data.

Since our study focuses on the impact of race on debt collection judgments, it is essential to determine if minorities are underrepresented in the data or if a specific type of minority borrower, such as high or low-income, is underrepresented. We next investigate whether race plays a role in the likelihood of matching originated mortgages from the HMDA database to our sample of credit bureau records. The regressions in Table \ref{table:2} assess the probability of successfully matching a loan based on borrower race. The coefficients on Black and Hispanic, and the interaction terms Black X log(Income) and Hispanic X log(Income) in column (1) (2) are insignificant. While there is a slight indication of selection bias through the combination of race and income for Black applicants seeking to refinance mortgages, we can control for variables from both databases, which mitigates concerns about selection bias.

Table \ref{table:3} provides summary statistics for our individual-level panel data starting in 2013Q2 and ending in 2017Q2.\footnote{While a larger sample of data is used to match credit report and HMDA data, we limit our sample for our regression analysis as judgment data isn't available in every quarter of the credit report data.} Similar to patterns identified in \citeN{butler2022racial}, Columns 1 and 2 indicate that individuals within the matched data set generally exhibit higher credit scores, are relatively younger, and have a higher likelihood of holding a mortgage compared to the average U.S. resident who has a credit history. These patterns are anticipated, as individuals must either obtain a new mortgage or refinance an existing one between 2010 and 2017 to qualify for inclusion in the matched panel. Columns 3–5 demonstrate that the White/Asian borrowers in the matched panel have higher credit scores and incomes in comparison to minority (Black/Hispanic) borrowers.

\input{Tables/tab_1}
\input{Tables/tab_2}
\input{Tables/tab_3}

\clearpage 

\subsection*{Match Validation}

To verify the accuracy of the credit bureau data, we plot the evolution of our matching variables from 2010-2017. We then juxtaposed these results with data obtained from the Home Mortgage Disclosure Act (HMDA), presenting our findings in Figure \ref{fig:experian_validation_part}. This figure comprises five panels: Panel (a) illustrates loan type trends, Panel (b) shows trends in purchaser types, Panel (c) details origination balances, Panel (d) depicts loan purpose trends, and Panel (e) focuses on match rates. In each panel, the left side represents credit bureau data, while the right side displays HMDA data.

Panels (a) through (c) reveal remarkably similar trends between the two data sources. In Panel (d), we observe a comparable trend with an increase in home purchase loans and a decrease in refinance loans. However, the credit bureau data shows a consistent 15-percentage point higher rate for purchase loans. This discrepancy is not a significant concern for two reasons. First, our ratio of home purchase to refinance mortgages aligns with that identified in \citeN{butler2022racial}. Considering their dataset is three times larger, our loan numbers are roughly one third in both categories. Second, to distinguish between transferred and refinanced first mortgages, we employed a criterion based on changes in the credit amount within the open first mortgage balance over the past two quarters. Importantly, refinancing does not always alter the loan amount; borrowers may refinance for better rates or terms without changing the principal, and we would not be able to capture those. 

Panel (e) shows the percent of observations we matched in the Credit Bureau and HMDA data respectively. The match rate in the Credit Bureau data remains relatively steady over time, consistently around 40\% until 2016-2017, when it increases to around 50\%. Similarly, the HMDA data shows a stable match rate, averaging about 0.1\% in most years, with a slight uptick to approximately 0.13-0.14\% in the later years. These trends indicate a general stability in the match rate across both datasets over the analyzed period.

\begin{figure}[!htbp]
\footnotesize
\centering
\begin{subfigure}[b]{0.95\textwidth}
\includegraphics[width=\textwidth]{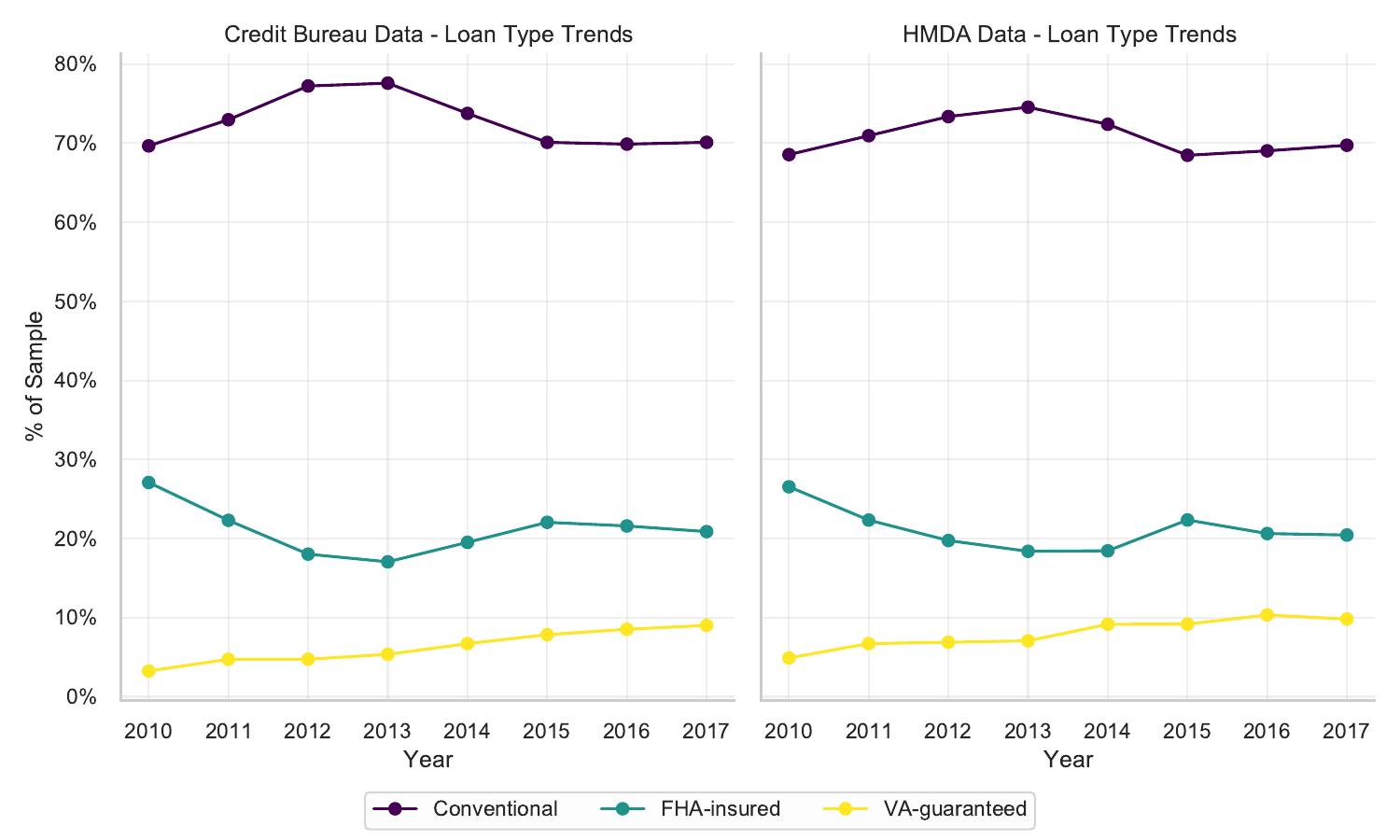}
\caption{}
\end{subfigure}

\begin{subfigure}[b]{0.95\textwidth}
\includegraphics[width=\textwidth]{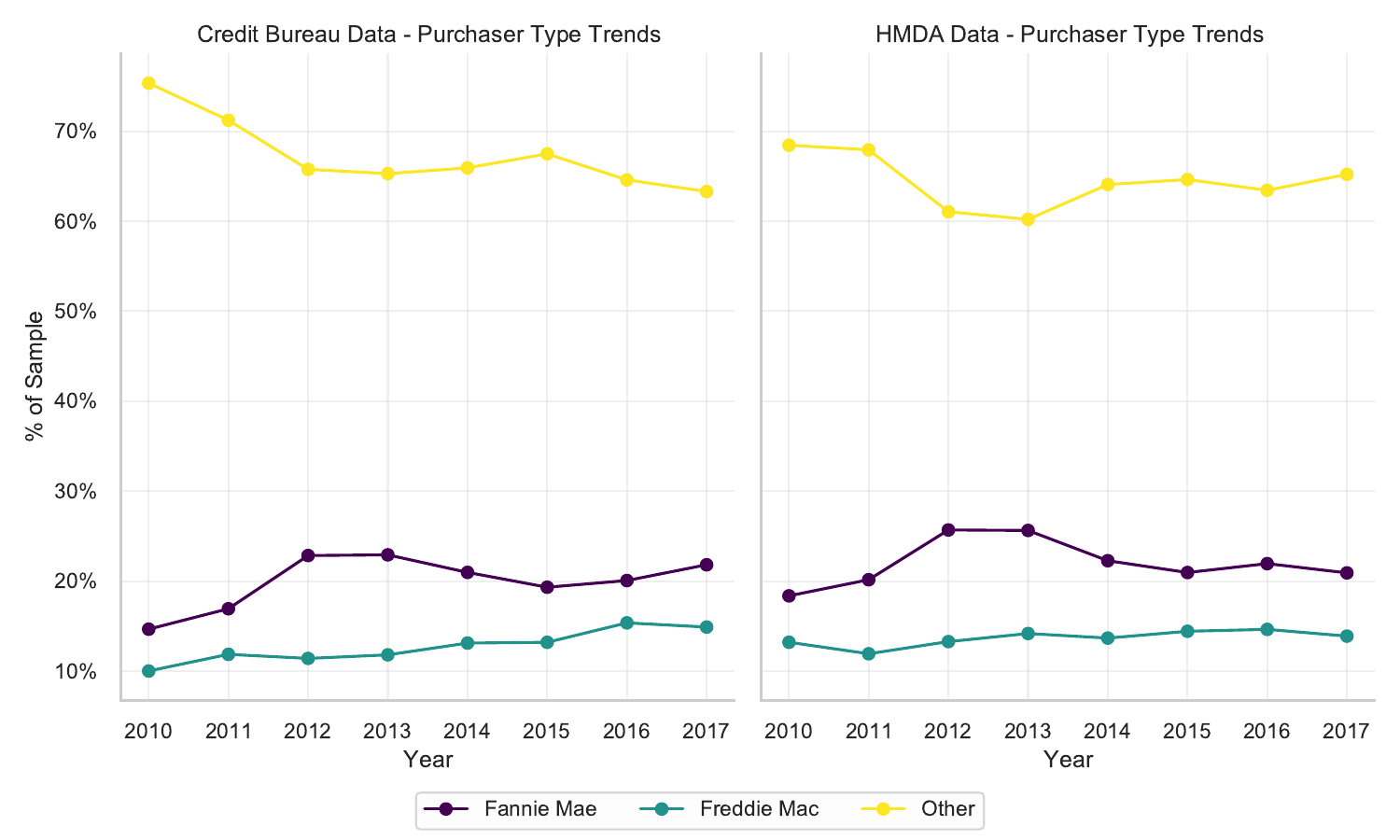}
\caption{}
\end{subfigure}
\caption{Credit Bureau \& HMDA trends (part 1).}\label{fig:experian_validation_part}
\end{figure}

\begin{figure}[!htbp]
\ContinuedFloat 
\footnotesize
\centering
\begin{subfigure}[b]{0.95\textwidth}
\includegraphics[width=\textwidth]{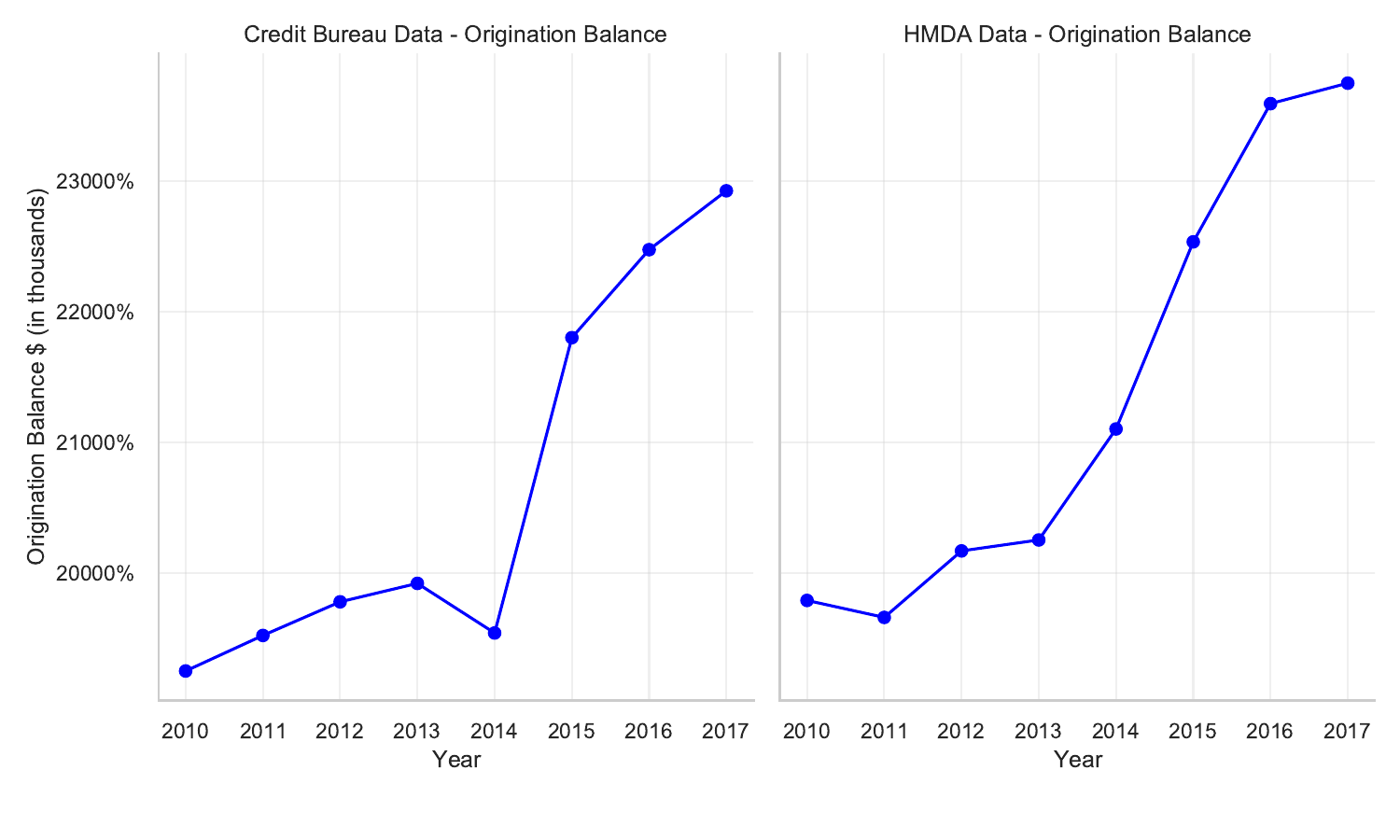}
\caption{}
\end{subfigure}

\begin{subfigure}[b]{0.95\textwidth}
\includegraphics[width=\textwidth]{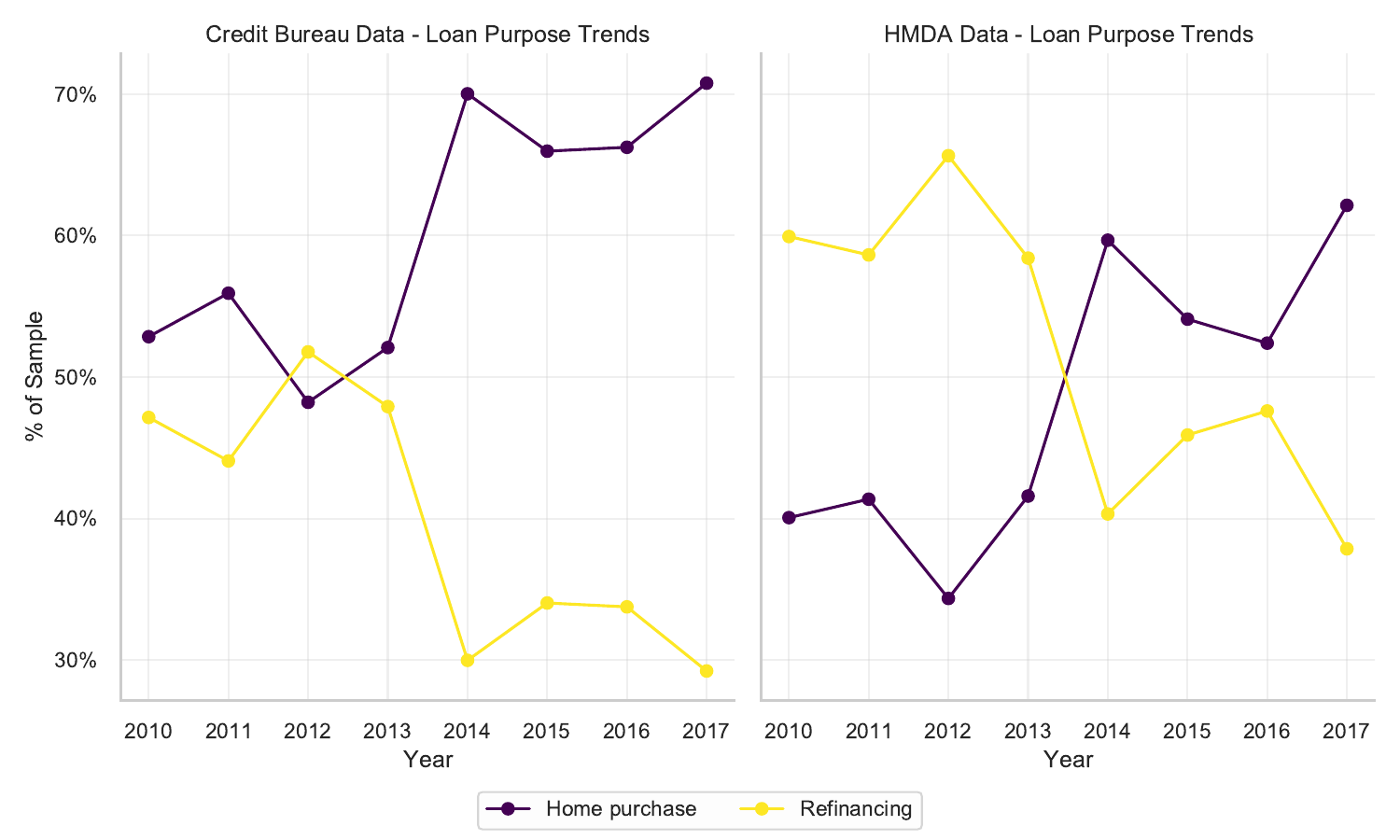}
\caption{}
\end{subfigure}
\caption{Credit Bureau \& HMDA trends (part 2).}\label{fig:experian_validation_part2}
\end{figure}

\begin{figure}[!htbp]
\ContinuedFloat 
\footnotesize
\centering
\begin{subfigure}[b]{0.95\textwidth}
\includegraphics[width=\textwidth]{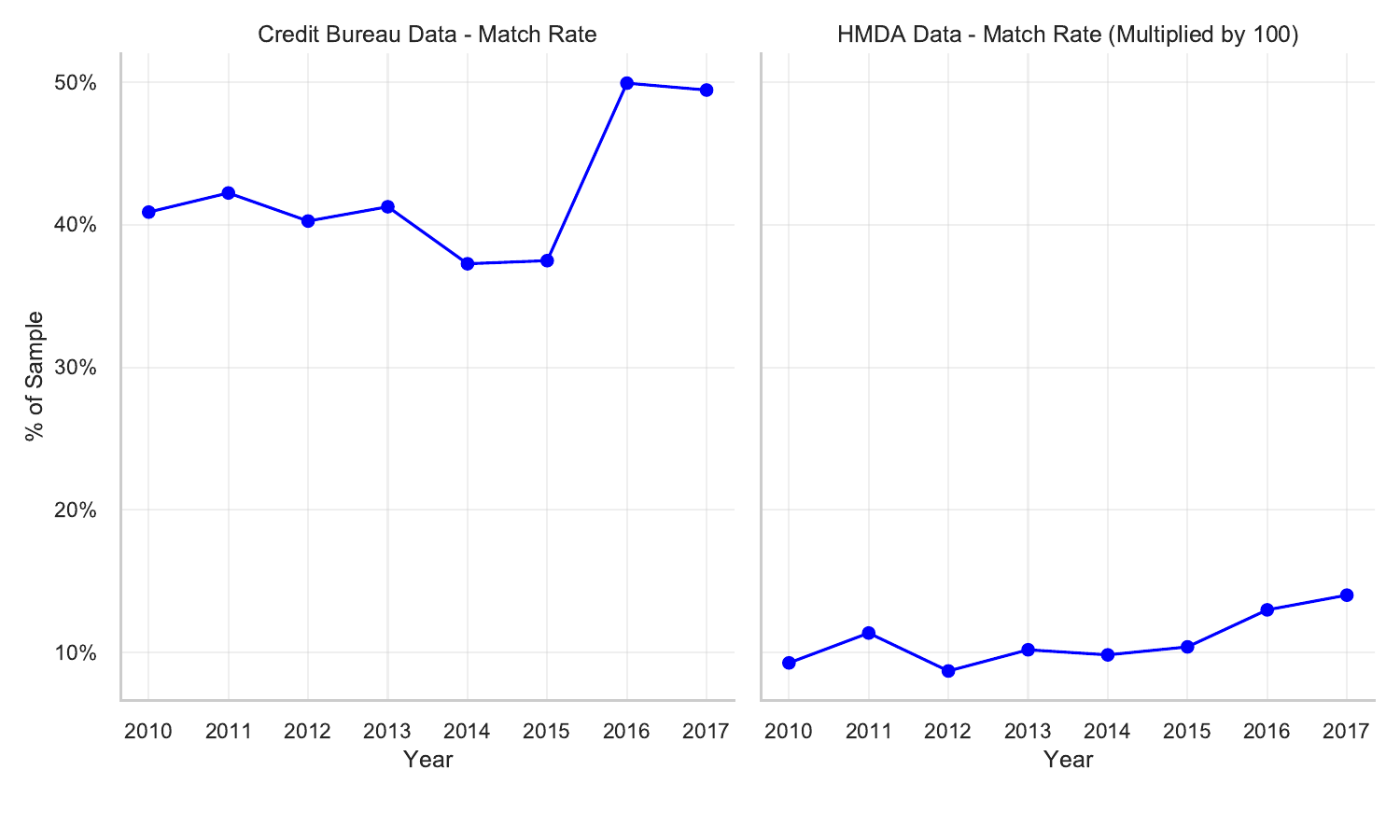}
\caption{}
\end{subfigure}

\caption{Credit Bureau \& HMDA trends (part 3).}\label{fig:experian_validation_part3}
\begin{flushleft}
\footnotesize Notes: Figure \ref{fig:experian_validation_part} compares Credit Bureau and HMDA data from 2010-2017. Panels (a) to (e) display trends in loan types, purchaser types, origination balances, loan purposes, and match rates, respectively. In each panel, the left side represents credit bureau data, while the right side displays HMDA data. The HMDA match rate in Panel (e) is multiplied by 100 for clearer trend visualization. Both data sources exhibit similar trends, particularly in Panels (a) to (c). Panel (d) shows a consistently higher rate of purchase loans in Credit Bureau data, similar to \citeN{butler2022racial}.
\end{flushleft}
\end{figure}

\subsection*{Credit Bureau Data Matching Variables}\label{appendixa}

We matched mortgages in the credit bureau data to HMDA data based on six characteristics:  ZCTA5, loan amount, loan purpose (purchase or refinancing), origination year, mortgage type (conventional, FHA, or VA), and if/to which quasi-government entity the loan is sold (purchaser type). 

To enhance the accuracy of matching correct balances, especially in refinancing cases, our credit bureau sample is restricted to individuals with a single first mortgage trade in the current quarter and at most one in the past two quarters.\footnote{This restriction increases the likelihood that the properties in question are owner-occupied when analyzing purchase transactions.} Utilizing the feature that records the total credit amount on open first-lien mortgages, we can accurately determine the loan amount. Additionally, with ZCTA5 directly observable in our data, two of the six matching variables required for our analysis are addressed.

For loan purpose determination, we analyze the total credit amount on open first mortgage trades, the time since most recent closure, transfer or refinancing of first-lien mortgages, the time elapsed since the most recent first-lien mortgage trade opening, and the number of outstanding first-lien mortgages. A mortgage is considered ongoing and not closed if there's an active credit balance in the current quarter, differing from the previous two quarters. This change in balance indicates that the transaction is a new financial activity, either a new purchase or a refinancing. For purchase identification, we look for loans opened within the past six months with no recorded open first-lien mortgages in the preceding two quarters. This confirms new purchases rather than extensions. For refinances, we identify mortgages with recent renegotiations or refinancing, evidenced by a mortgage closure, transfer, or refinancing in the last six months and an outstanding first-lien mortgage in the preceding two quarters.

The origination year for refinance mortgages is determined by subtracting the time since the mortgage's most recent closure, transfer, or refinancing from the current date. For purchase mortgages, it's calculated from the time elapsed since the opening of the latest first-lien mortgage trade.

Loan types are classified using a specific variable in the credit bureau dataset that categorizes the industry type of the most recently opened first mortgage in the past six months as VA, FHA, or conventional.

Purchaser type is determined by comparing the balances of open first mortgage trades for Fannie Mae and Freddie Mac over the last six months. A mortgage is classified based on which entity's balance is higher, falling into the 'Other' category if neither. Our data does not include cases where one balance is greater than the other while the lesser balance is not zero. This category also includes loans sold to private institutions, as our data does not distinguish these sales.

%% file: Tables/sample_restrictions_hmda.tex
\begin{sidewaystable}[htbp]\centering
\begin{threeparttable}
\small
\caption{Itemized Sample Restrictions: Credit Bureau and HMDA Data}\label{tab:sample_restrictions_experian}
\begin{tabular}{lc} \hline \hline
& \\[\dimexpr-\normalbaselineskip+3pt]
\underline{Credit Bureau Data [2010-2017]} & \\
& \\[\dimexpr-\normalbaselineskip+3pt]
Removal Reason &  \# of Individuals  \\ \hline 
& \\[\dimexpr-\normalbaselineskip+3pt]
Initial Data & 769,423 \\
No Mortgage & 310,542 \\
First Lien  Mortgage Balance = Joint Mortgage Balance & 290,481 \\
Non-HMDA zip codes & 267,042 \\
More than one first mortgage (past two quarters and current) & 127,891 \\
First mortgage balance is zero or it is not different from the past two quarters & 79,226 \\
Foreclosed balances & 77,481 \\
Refinance or Purchase Activity over 6 months ago & 44,192 \\ \hline
& \\
\underline{HMDA Data [2010-2017]} & \\
& \\[\dimexpr-\normalbaselineskip+3pt]
Removal Reason &  \# of Loans  \\ \hline 
& \\[\dimexpr-\normalbaselineskip+3pt]
Initial Data & 62,610,270 \\
Missing MSA& 55,412,483 \\
Home improvement loans & 52,565,523 \\
Joint loans & 25,909,081 \\
Non-first lien loans & 25,501,830 \\
Missing geographic information or outside of the 50 States + D.C. & 25,500,339 \\
Non-owner occupied loans & 22,565,758 \\
Duplicate observations along the match variables at the census tract level,20,250,348 \\
Tracts with no associated ZCTAs& 19,334,339 \\
Tract-ZCTA5 Expansion & 35,446,686 \\
Duplicates along match variables at the ZCTA5 level & 27,413,377 \\ \hline \hline 

\end{tabular}
\begin{tablenotes}
\footnotesize
\item Notes:  This table documents the sample restrictions applied to the credit bureau and HMDA datasets. The final credit bureau data contains 50,166 observations of 44,192 individuals. The final HMDA dataset, after deduplication at the ZCTA level, contained 27,413,377 observations, which included 15,763,550 of the original census tract level loans.
\end{tablenotes}
\end{threeparttable}
\end{sidewaystable}

%% file: Tables/tab_1.tex
\begin{table}[!ht]
\footnotesize
    \centering
    \begin{threeparttable}
        \caption{Descriptive Statistics of Matching Data from Credit Bureau with HMDA}  
        \label{table:1}
        \begin{tabular}{l c c c c c}  \hline \hline
        & & & & &  \\[\dimexpr-\normalbaselineskip+2pt]
            \multicolumn{6}{l}{\textbf{Panel A: Match rate}} \\ \hline
             & & & & &  \\[\dimexpr-\normalbaselineskip+2pt]
            & Credit bureau   & Matched to  & Match rate (\%) & & \\
            &  sample &  HMDA &  & & \\
Home Purchase Mortgages         & 31,411                         & 12,761                                  & 40.63                                   \\ 
Refinance Loans                 & 19,919                         & 9,320                                   & 46.79                                   \\
All Loans                       & 51,330                         & 22,081                                  & 43.02                                   \\
\hline
            & & & & &  \\[\dimexpr-\normalbaselineskip+2pt]
            \multicolumn{6}{l}{\textbf{Panel B: Home purchase mortgages}} \\ \hline
            & & & & &  \\[\dimexpr-\normalbaselineskip+2pt]
             & Credit bureau  & Matched to  & Unmatched & \multicolumn{2}{c}{\underline{Matched vs. Unmatched}} \\
             &  sample &  HMDA &  & Norm. diff. & t-stat  \\
            \underline{Match criteria} & & & & & \\
            & & & & & \\[\dimexpr-\normalbaselineskip+2pt]
Conventional                   & 0.640                          & 0.591                    & 0.674              & -0.17                & -15.03         \\
FHA-insured                    & 0.285                          & 0.327                    & 0.257              & 0.16                 & 13.62          \\
VA-guaranteed                  & 0.075                          & 0.082                    & 0.070              & 0.05                 & 4.03           \\ 
Fannie Mae                     & 0.207                          & 0.182                    & 0.223              & -0.10                & -8.78          \\ 
Freddie Mac                    & 0.130                          & 0.103                    & 0.148              & -0.14                & -11.81         \\ 
Loan amount                    & 202,432                    & 193,610             & 208,468        & -0.08                & -7.15          \\ 
            & & & & & \\[\dimexpr-\normalbaselineskip+2pt]
            \underline{Non-match characteristics} & & & & & \\
            & & & & & \\[\dimexpr-\normalbaselineskip+2pt]
Credit score t-1               & 716                      & 710                & 720            & -0.15                & -12.02         \\ 
Age                            & 40.7                        & 39.4                   & 41.5             & -0.15                & -12.26         \\ 
Have mortgage t-1              & 0.099                          & 0.064                    & 0.123              & -0.20                & -17.17         \\ 
Total debt t-1                 & 29,933                    & 26,043               & 32,472        & -0.10                & -8.08          \\ 
Past due debt t-1              & 137                      & 155                  & 125            & 0.01                 & 0.86           \\

 \hline
            Observations &  31,411 &  12,761 & 18,650 &  &  \\ \hline \hline
            & & & & & \\[\dimexpr-\normalbaselineskip+2pt]
            \multicolumn{6}{l}{\textbf{Panel C: Refinance loans}} \\  \hline
             & Credit bureau  & Matched to & Unmatched & \multicolumn{2}{c}{\underline{Matched vs. Unmatched}} \\
              &  sample & HMDA &  &  Norm. diff & t-stat   \\
             \underline{Match criteria} & & & & & \\
             & & & & & \\[\dimexpr-\normalbaselineskip+2pt]
Conventional                   & 0.861                          & 0.842                    & 0.877              & -0.10                & -7.17           \\ 
FHA-insured                    & 0.087                          & 0.102                    & 0.075              & 0.10                 & 6.74            \\ 
VA-guaranteed                  & 0.052                          & 0.056                    & 0.048              & 0.04                 & 2.60            \\ 
Fannie Mae                     & 0.202                          & 0.213                    & 0.192              & 0.05                 & 3.70            \\ 
Freddie Mac                    & 0.131                          & 0.128                    & 0.134              & -0.02                & -1.26           \\ 
Loan amount                    & 218,933                   & 202,233              & 233,619       & -0.17                & -12.33          \\ 

            & & & & & \\[\dimexpr-\normalbaselineskip+2pt]
            \underline{Non-match characteristics} & & & & & \\
            & & & & & \\[\dimexpr-\normalbaselineskip+2pt]
Credit score t-1               & 739                        & 739                 & 739           & -0.00                & -0.16           \\ 
Age                            & 49.3                         & 49.6                   & 49.1             & 0.04                 & 2.96            \\ 
Have mortgage t-1              & 1                          & 1                    & 1              & 0                  & 0             \\ 
Total debt t-1                 & 240,201                    & 224,718              & 253,816        & -0.15                & -10.70          \\ 
Past due debt t-1              & 1,039                      & 540                  & 1,477          & -0.05                & -3.51           \\ 

 \hline 
            Observations    & 19,919 & 9,320 & 10,599 &  & \\
                \hline \hline
\end{tabular}
\begin{tablenotes}
\footnotesize
\item Notes: This table provides an overview of the credit bureau and the Home Mortgage Disclosure Act (HMDA) data matching. The initial sample of credit bureau mortgages comprises both home purchase mortgages and refinance loans issued between 2010 and 2017. Loan applicants must apply individually (excluding joint applications), reside within a metropolitan statistical area after loan origination in one of the 50 states (including D.C.), and have the mortgage as their sole first-lien mortgage. The matching process is described in Section 3.1. Panel A illustrates the success rate of the matching methodology. Panel B summarizes loan and borrower features for home purchase mortgages in the credit bureau data, the subset successfully matched to HMDA, and the unmatched loans. The final two columns display the normalized difference and the outcome of a t-test comparing the mean of the matched sample to the mean of the unmatched sample. Panel C offers comparable summary statistics for refinance loans. 
\end{tablenotes}
\end{threeparttable}
\end{table}
  

%% file: Tables/tab_2.tex
\begin{table}[!ht]
\centering
\def\sym#1{\ifmmode^{#1}\else\(^{#1}\)\fi}
\small
\begin{threeparttable}
\caption{Does borrower race affect the Credit Bureau/HMDA match?}
\label{table:2}
\begin{tabular}{lccc}
\hline \hline
& & & \\[\dimexpr-\normalbaselineskip+2pt]
& \multicolumn{1}{c}{Full sample} & \multicolumn{1}{c}{Home purchase mortgages} & \multicolumn{1}{c}{Refinance loans} \\
& (1) & (2) & (3) \\
\hline
& & & \\[\dimexpr-\normalbaselineskip+2pt]
\textbf{Match criteria} & & & \\
& & & \\[\dimexpr-\normalbaselineskip+2pt]
FHA loan & 0.0070\sym{***} & 0.0369\sym{***} & -0.0567\sym{***} \\
 & (0.0023) & (0.0032) & (0.0034) \\
VA loan & -0.0135\sym{***} & 0.0110\sym{**} & -0.0524\sym{***} \\
 & (0.0031) & (0.0044) & (0.0044) \\
Purchased by Fannie Mae & -0.0231\sym{***} & 0.0321\sym{***} & -0.0580\sym{***} \\
 & (0.0020) & (0.0036) & (0.0025) \\
Purchased by Freddie Mac & -0.0316\sym{***} & 0.0214\sym{***} & -0.0637\sym{***} \\
 & (0.0023) & (0.0044) & (0.0028) \\
log(Loan amount) & -0.0173\sym{***} & -0.0366\sym{***} & -0.0036\sym{*} \\
 & (0.0015) & (0.0027) & (0.0018) \\
& & & \\[\dimexpr-\normalbaselineskip+2pt]
 \textbf{Non-match characteristics} & & & \\
& & & \\[\dimexpr-\normalbaselineskip+2pt]
Black & -0.0024 & 0.0124 & 0.0109 \\
 & (0.0070) & (0.0536) & (0.0072) \\
Hispanic & -0.0083 & 0.0283 & -0.0036 \\
 & (0.0076) & (0.0459) & (0.0077) \\
Black X log(Income) & -0.0003 & -0.0018 & -0.0019\sym{**} \\
 & (0.0007) & (0.0048) & (0.0007) \\
Hispanic X log(Income) & 0.0010 & -0.0028 & 0.0003 \\
 & (0.0007) & (0.0042) & (0.0008) \\
log(Income) & 0.0031\sym{***} & -0.0048\sym{***} & 0.0006\sym{*} \\
 & (0.0003) & (0.0012) & (0.0003) \\ \hline 
& & & \\[\dimexpr-\normalbaselineskip+2pt]
Census tract-by-year FE & Yes & Yes & Yes \\ 
& & & \\[\dimexpr-\normalbaselineskip+2pt]
R-squared & 0.0027 & 0.0042 & 0.0092 \\
Observations & 20,250,348    & 9,706,546    & 10,543,802    \\ \hline  \hline
\end{tabular}
\begin{tablenotes}
\footnotesize
\item Notes: This table presents regression analyses investigating the factors influencing the matching of a mortgage from the Home Mortgage Disclosure Act (HMDA) data to a credit bureau record. The sample includes all first-lien home purchase mortgages and refinance loans for owner-occupied properties in metropolitan statistical areas between 2010 and 2017 with reported income. Only single-applicant loans are considered, excluding joint applications. The dependent variable is a binary indicator representing a successful HMDA mortgage match to a credit bureau record, while the independent variables are loan and borrower attributes from the HMDA data. Columns 1, 2, and 3 display the results for the entire sample, home purchase mortgages, and refinance loans, respectively. The coefficients are reported in percentage point units. Standard errors are clustered by census tract-year. \sym{*} \(p<0.10\), \sym{**} \(p<0.05\), \sym{\sym{***}} \(p<0.01\).
\end{tablenotes}
\end{threeparttable}
\end{table}

%% file: Tables/tab_3.tex
\begin{table}[!ht]
\small
\centering
\begin{threeparttable}
\caption{Summary Statistics for the Credit Bureau/HMDA Matched Panel}\label{table:3}
\begin{tabular}{lcccccc}
\hline \hline
& & & & &  \\[\dimexpr-\normalbaselineskip+3pt]
 & \multicolumn{1}{p{1.5cm}}{\centering Full credit sample} & \multicolumn{1}{p{1.5cm}}{\centering Matched sample} & \multicolumn{1}{p{1.5cm}}{\centering White/Asian} & \multicolumn{1}{p{1.5cm}}{\centering Black} & \multicolumn{1}{p{1.5cm}}{\centering Hispanic}\\
\hline
& & & & & \\[\dimexpr-\normalbaselineskip+3pt]
Age & 50.86 & 44.94 & 45.21 & 45.00 & 42.68 \\
 & (19.13) & (14.25) & (14.37) & (14.26) & (13.02)\\
Household income & 74.80 & 93.25 & 96.43 & 77.45 & 79.28 \\
 & (43.72) & (51.49) & (52.62) & (41.24) & (44.04)\\
Credit score$_{t-1}$ & 682.66 & 716.92 & 723.31 & 677.29 & 694.14 \\
 & (103.49) & (81.23) & (79.03) & (87.61) & (81.52)\\
Have mortgage$_{t-1}$ & 0.268 & 0.687 & 0.699 & 0.641 & 0.627 \\
 & (0.443) & (0.464) & (0.459) & (0.480) & (0.484)\\
Total debt$_{t-1}$ & 64,184 & 154,224 & 159,179 & 128,788 & 132,477 \\
 & (151,233) & (169,753) & (175,425) & (122,437) & (148,105)\\
30 DPD debt$_{t-1}$ & 776.72 & 1,087.03 & 962.12 & 2,031.44 & 1,402.83 \\
 & (13,062.49) & (14,750.96) & (14,439.15) & (17,405.45) & (15,066.73)\\
60 DPD debt$_{t-1}$ & 323.00 & 391.49 & 334.74 & 844.95 & 516.25 \\
 & (9,031.04) & (8,324.76) & (7,912.57) & (11,289.45) & (8,957.47)\\
90+ DPD debt$_{t-1}$ & 1,892.81 & 1,205.09 & 1,073.66 & 2,577.32 & 1,246.79 \\
 & (22,307.79) & (14,087.02) & (13,660.40) & (18,740.64) & (13,350.45)\\
90+ DPD on revolving debt$_{t-1}$ & 153.49 & 115.30 & 114.81 & 137.01 & 102.70 \\
 & (4,963.97) & (2,016.56) & (2,095.42) & (1,800.13) & (1,413.69)\\
90+ DPD on installment debt$_{t-1}$ & 396.27 & 189.86 & 151.62 & 584.17 & 205.82 \\
 & (6,653.93) & (3,419.63) & (3,141.07) & (5,929.77) & (2,962.45)\\
Collections debt$_{t-1}$ & 609.45 & 208.53 & 167.75 & 406.89 & 395.57 \\
 & (3,955.72) & (1,720.72) & (1,627.30) & (1,673.76) & (2,368.18)\\
Public record bankruptcies$_{t-1}$ & 0.0489 & 0.0527 & 0.0486 & 0.0857 & 0.0618 \\
 & (0.2354) & (0.2344) & (0.2222) & (0.3202) & (0.2531)\\
Public record judgments$_{t-1}$ & 0.0136 & 0.0080 & 0.0069 & 0.0179 & 0.0093 \\
 & (0.1378) & (0.0995) & (0.0937) & (0.1450) & (0.1030)\\
\hline
 & & & & & \\[\dimexpr-\normalbaselineskip+3pt]
Individuals  & 760,680 & 19,061  & 15,713  & 1,445  & 1,903 \\ \hline \hline
\end{tabular}
\begin{tablenotes}
\item \footnotesize{This table provides descriptive statistics for the Credit Bureau/HMDA matched panel. Column 1 outlines the sample means and standard deviations for the entire credit bureau dataset; column 2 portrays these statistics for the Credit Bureau/HMDA matched panel; and columns 3–6 display these statistics for the White/Asian, Black, and Hispanic borrowers within the matched dataset, respectively. The sample spans 2013Q2 to 2017Q2. }
\end{tablenotes}
\end{threeparttable}
\end{table}